# Can Small and Reasoning Large Language Models Score Journal Articles for Research Quality and Do Averaging and Few-shot Help?[1]

Mike Thelwall, School of Information, Journalism and Communication, University of Sheffield, UK. https://orcid.org/0000-0001-6065-205X m.a.thelwall@sheffield.ac.uk
Ehsan Mohammadi, School of Information Science, University of South Carolina, 1501 Greene St, Columbia, SC 29208, USA. https://orcid.org/0000-0003-3481-6991

Previous research has shown that journal article quality ratings from the cloud based Large Language Model (LLM) families ChatGPT and Gemini and the medium sized open weights LLM Gemma3 27b correlate moderately with expert research quality scores. This article assesses whether other medium sized LLMs, smaller LLMs, and reasoning models have similar abilities. This is tested with Gemma3 variants, Llama4 Scout, Qwen3, Magistral Small and DeepSeek R1 on a dataset of 2,780 medical, health and life science papers in 6 fields, with two different gold standards, one novel. Few-shot and score averaging approaches are also evaluated. The results suggest that medium-sized LLMs have similar performance to ChatGPT 4o-mini and Gemini 2.0 Flash, but that 1b parameters may often, and 4b sometimes, be too few. Reasoning models did not have a clear advantage. Moreover, averaging scores from multiple identical queries seems to be a universally successful strategy, and there is weak evidence that few-shot prompts (four examples) tend to help. Overall, the results show, for the first time, that smaller LLMs >4b have a substantial capability to rate journal articles for research quality, especially if score averaging is used, but that reasoning does not give an advantage for this task; it is therefore not recommended because it is slow. The use of LLMs to support research evaluation is now more credible since multiple variants have a similar ability, including many that can be deployed offline in a secure environment without substantial computing resources.

## Introduction

Whilst citation-based indicators have been the primary source of quantitative support for expert judgements about research quality or impact, Large Language Models (LLMs) offer an alternative. Using a departmental-level proxy for research quality scores (discussed below), there is now substantial evidence that the large closed-source public LLMs ChatGPT 4o, ChatGPT 4o-mini (Thelwall & Yaghi 2025; Thelwall et al., 2025), and Gemini 1.5 Flash (Thelwall, 2025b) produce values that correlate with expert judgement at levels comparable with citation-based indicators, with ChatGPT 4o and 4o-mini exceeding them in most fields (Thelwall, 2025e). There is also evidence of moderate correlations with the same research quality proxy for the "open weights" (downloadable) LLM Gemma3 27b (Thelwall, 2025a). ChatGPT 4o and 4o-mini scores also correlate positively with direct expert research quality scores for one small sample (Thelwall, 2024, 2025c). Whilst two open weights LLMs, Qwen2.5-72b and Llama3, have been compared with ChatGPT 4o-mini against a proxy quality rating for 15,000 biomedical articles, only a weak ability was found (Cohen's kappa of 0.059, also trying few-shot and fine tuning: Wu et al., 2025). Thus, is not known whether open weights LLMs other than Gemma3 27b, and particularly smaller LLMs, can rate journal articles, whether reasoning models (see below) have an advantage, and whether the few-shot prompting technique helps.

[1] Thelwall, M. & Mohammadi, E. (2026). Can small and reasoning Large Language Models score journal articles for research quality and do averaging and few-shot help. Scientometrics.

Although a larger LLM is generally better (Sardana et al., 2023) and size can be critical for some tasks (Berti et al., 2025), the relationship between LLM size and research quality scoring performance is unknown. The LLM size-performance relationship issue is obscured by LLMs having different architectures and different capabilities (e.g., multilingual, multimodal). Size calculations are also complicated by LLMs having a "full" size at which they were initially built and versions that may retain some benefits of the original size. Quantised versions of an LLM are smaller due to reducing the accuracy of the parameters in the model (Xiao et al., 2023). In contrast, checkpoint/snapshot versions of an LLM have different numbers of parameters but presumably similar overall architectures and training data. Differently again, "knowledge distillation" versions of LLMs use training to generate a smaller LLM from a larger one through a combination of training on data and learning from the parameters of the bigger model. The current paper only uses checkpoint or distilled versions of larger LLMs (when known).

Reasoning or thinking models are LLMs trained to respond to prompts through multiple separate but interconnected steps (e.g., self-generated chain-of-thought) and to revise earlier steps in the light of later evidence (Venhoff et al., 2025). Reasoning models may also leverage external sources, such as web searches or a code execution environment (Plaat et al., 2025). They were first introduced in September 2024 with the OpenAI o1 preview. Subsequent models with the same approach include Alibaba's Qwen-based QwQ in November 2024, DeepSeek R1 with web search enhanced reasoning (January 2025), OpenAI o3 (February 2025), Anthropic's Claude 3.7 Sonnet (February 2025), Qwen 3 (April 2025), and Mistral's Magistral (June 2025). The performance of these models on research quality evaluation tasks has not been explored yet.

Based on the above discussion, three main research questions are addressed here. The fourth research question takes advantage of the multiple evaluations reported in the paper to assess whether the commonly used averaging strategy for scores is universally effective.

- RQ1: *Do other medium sized (approximately 30b) LLMs score academic research with a similar performance (correlation with expert scores) to Gemma2 27b, and does reasoning help?* A similar capability for a range of models would suggest that it is core to medium sized LLMs – and hence likely to persist -rather than a byproduct of one architecture.
- RQ2: *What is the relationship between LLM size (number of parameters or file size) and journal article research quality scoring ability? Does this ability emerge at a given size (e.g., Wei et al., 2022), does it scale more smoothly with size, does it peak at a given size or something else?* The answer may vary between LLMs, but it may be possible to give a baseline expectation based on one.
- RQ3: *Can the few-shot prompting strategy improve an LLM's ability to evaluate research papers?* One previous study has suggested that this is possible for a related task (Wu et al., 2025).
- RQ4: *Is averaging scores across multiple iterations of identical prompts universally effective at increasing score alignment with expert scores?* Previous studies have shown that scoring an article multiple times and taking the mean score gives more useful information (e.g., Thelwall, 2024) but it is not known whether this is universally true.

This study does not investigate all possible strategies for improving scores because there are too many. For this reason, many LLMs are not tested, and particularly older ones, and non-generative transformer models (e.g., BERT-type). In addition, parameter variations are not tested, and neither are variations in prompts or inputs. Fine-tuning and agent-based designs are also ignored. All might produce improved results, but they are out of scope here.

# Methods

The research design was to create a large dataset of articles from multiple fields with expert scores, then apply the different LLMs and prompting strategies to score them, correlating the human and AI scores. Previous research has shown that LLM accuracy is low because LLM scores tend to cluster at a particular value, so the most important property is rank correlation: the extent to which the order of the articles is similar between the expert scores and the LLM scores.

## *Data*

Except for a few publications with short-term access to private data, previous studies using research quality for journal articles have either used small datasets of under 100 articles with individual research quality scores from a single field (Thelwall, 2024, 2025c), a medium sized dataset from a single broad field with a research quality proxy (Wu et al., 2025), or large scale science wide data or using departmental mean scores as a proxy for individual article scores in the UK's Research Excellence Framework (REF) 2021 (Pölönen et al., 2023; Thelwall & Kurt, 2025; Thelwall & Yaghi, 2025; Thelwall & Yang, 2025) or previous versions (Pride & Knoth, 2018). The current study initially followed the commonly used departmental mean score approach above, although the largest REF2021 dataset was impractical because LLMs can be expensive to run and particularly, as in the current paper, when several LLMs are to be compared against each other, multiple iterations are needed, and reasoning models are included. Thus, the current paper focuses on six fields in one broad area, the health and life sciences (Main Panel A of UK REF2021), and samples of 500 papers from these fields to keep the sizes manageable. This therefore follows an established proxy strategy for one gold standard: instead of using the (unknown) individual article quality scores, using the (known) departmental average quality scores for each of the department's articles. The effect of this on correlations is discussed below. Individual article-level quality scores were also created for this set as a second gold standard for triangulation, as explained below.

Main Panel A in REF2021 is an organisational unit including the following six field-based Units of Assessment (UoAs): 1 Clinical Medicine; 2 Public Health, Health Services and Primary Care; 3 Allied Health Professions, Dentistry, Nursing and Pharmacy; 4 Psychology, Psychiatry and Neuroscience; 5 Biological Sciences; 6 Agriculture, Food and Veterinary Sciences. Although there are 34 UoAs in REF2021 these six account for 41% of all journal articles submitted to it. Each includes over 3000 articles, but a random number generator was used to select 500 from each, after excluding articles without DOIs (needed for checking uniqueness) and without short or missing abstracts (i.e., excluding articles with the 10% shortest abstracts that are often short-form contributions). Articles can be submitted to different UoAs by different institutions so there was a small degree of overlap, with 2780 unique articles altogether.

The heart of REF2021 involves 1,120 experts (mostly senior professors) reading the journal articles submitted (and other outputs) and assigning them a score on the following scale (REF2021, 2021):

- 1* Nationally recognised
- 2* Internationally recognised
- 3* Internationally excellent
- 4* World leading

The scoring is based on a qualitative assessment of the rigour, originality and significance of the submitted work. In Main Panel A, the assessors could consult a limited set of bibliometric indicators for the older papers, showing their percentile rank for citations within the fields assigned to them by the Web of Science. These indicators had a minor role compared to the qualitative judgements of the two experts assigned to them (Wilsdon et al., 2015). Individual article scores were ratified by UoA panels and norm referenced across the entire REF. The scores direct seven years of the UK's block research grants to academic institutions and are also important for departmental and university reputations.

Unfortunately, the REF2021 individual article scores have been deleted and only departmental aggregate scores released (technically, only "submission" aggregate scores were released but these can be thought of as broadly like a department or school in scope). These aggregate scores (which consist of the percentage of papers assigned each of the quality scores) were used as a gold standard for the current paper (as for previous studies) by assigning each article the mean score of the scores of all papers from submitting department. For example, if a department had 90% 4* scores and 10% 3* scores then this procedure would assign 3.9* to all articles from that department. If the same paper had been submitted by multiple departments, the mean of the department mean scores was used instead (occasionally, article scores varied between UoAs).

The departmental mean score proxy "works" in the sense that, in the absence of bias, within a UoA, the higher the correlation with the proxy, the higher the correlation with the underlying data. The use of a proxy thus serves to dampen the correlation strength, so it does not reveal the underlying correlation strength but allows comparisons between LLMs and between approaches to help identify the best one. Because of the damping effect, it lacks the statistical power of the raw scores and will tend to underestimate the correlation strengths. It can also give misleading information if there are departmental-level biases, such as a LLM tending to favour the research from one department over that from another for reasons other than quality, such as writing style or majority theoretical/methodological orientation.

For triangulation purposes, because of the potential for departmental-level biases in LLM scores, a second gold standard was used, of research quality scores assigned to each of the 2,780 articles by the first author of the current article. These scores were assigned before any of the testing in the current article and without any conscious knowledge of any LLM scores assigned to these articles prior to the current study. The scores were assigned individually within UoAs on a nine-point scale following the REF2021 quality criteria (REF2021, 2019) and then norm referenced to give similar average scores to the averages given in the REF2021 UoA for journal articles. Articles in multiple UoAs were only given a single score, however, averaging the scores from the individual UoAs. This gold standard has the advantage that the scores are individual to articles rather than a proxy, but the disadvantage that the first author is not an expert in any of the six UoAs (although he has published in four of them: 2, 3, 4, and 5). In mitigation of the latter point, research quality assessment in the life and health sciences is more straightforward than in most other fields because it mostly lacks abstract theories (unlike many social sciences, and some arts and humanities), usually does not focus on mathematical elaborations (unlike much physics and mathematical sciences and some chemistry, engineering, and computer science), and the importance of the work seems often to be transparent from the abstract (unlike much mathematical and physical sciences, arts and humanities). Although it would have been preferable to get multiple people to classify these articles, this was impractical because it is a huge task that it does not seem fair or reasonable to ask of other researchers. This seems reasonable in the context of there being two independent gold standards, so results appearing in both are less likely to be due to bias.

### *Few-shot articles*

For the few-shot prompting approach the user prompt for an LLM must contain example inputs with recommended outputs. In the current paper, the few-shot approach is not a good match because the key output is a score, but this is typically produced by the LLM only as part of a longer report of several paragraphs. Since expert reports are never the same, there is no perfect "correct" answer for the few-shot strategy. Moreover, there is not a standard number of examples to use. Thus, any few-shot strategy must be a compromise.

The first decision was the number of examples to include within the prompt to guide the LLM about how to conduct the task: the "few" in few-shot. There are at least three reasonable strategies:

1. Two: a 1* article and a 4* article to indicate the score range.
2. Two: a 3* article and a 4* article since these are the most common scores.
3. Four: 1*, 2* 3* and 4* articles to indicate all levels.

The third approach was chosen because this seems to be the most natural strategy. Differentiating between the common 3* and 4* categories would not be helped much by strategy 1. Moreover, identifying rarer low-quality articles is important, a drawback for strategy 2. Whilst strategy 3 might be overcomplicated in the sense of producing longer prompts, they stay well within the context window of the main models used, so this does not seem to be a large risk. A previous study has included multiple articles from each category (five for each of the three levels) (Wu et al., 2025) but few-shot usually includes fewer examples so this approach was not attempted. There also seem to be three reasonable approaches for generating the few-shot examples.

A. Randomly select articles from the 2780 set with the author-assigned scores at the requisite levels.
B. Select four articles from each UoA set of 500 and use them for the remaining 496, reducing the effective sample size.
C. Select four articles outside the 2780 for each UoA.

The first two strategies were not used to protect against any of the platforms remembering and/or learning from user inputs, so using out-of-set articles is a safeguard against accidental leaking of the "correct" scores (including due to human error). Whilst two of the cloud-based LLMs promised not to incorporate any submitted data for training, they might still conceivably use them for short term responses from the submitting users (e.g., query caching). A disadvantage of Strategy C is that the choice of article at each level is subjective, and a mistake could influence the power of the few-shot strategy. Thus, a modified strategy "Cx2" was used: find two articles at each star level in each UoA (so 2 x 6 x 4 = 48 articles) and select one of the two articles at random at each star level for the relevant UoA when generating a few-shot prompt within a UoA.

The few-shot prompts were regenerated each time when needed, so if an article was submitted five times for a UoA few-shot experiment then it would usually have five different combinations of 1*, 2*, 3*, and 4* articles (but of course with a maximum of two different articles overall at each of the star levels).

### *Models*

The models used in the experiments are briefly introduced here to give context (Table 1). Unfortunately, the choice of models is partly pragmatic in the sense that larger open weights models are more expensive to run (e.g., needing more VRAM and more GPUs or longer), excluding most models larger than 60GB.

There are some general types of architecture for GenAI LLMs. The original LLMs had a "dense" architecture, with the entire model being applied to each input token in the prompt. In contrast, in "sparse" architectures, a subset of the model is applied to each input token, speeding the processing and reducing the hardware requirements. The Mixture of Experts (MoE) architecture is a sparse approach where the network is split into specialized subnetworks, with a gating mechanism directing each token to one or more experts for processing. A multimodal LLM is likely to use the MoE architecture, such as with separate sets of experts for text and vision, perhaps with additional fusion layers to combine information from different types of experts. Multilingual LLMs do not tend to be built around language-specialist MoE experts, but some experts may become naturally specialized in one or a few languages. Information seems more likely to transfer between languages when there are fewer language-specialist experts.

Table 1. Properties of the LLMs compared in this article, when known.

| Model | Date | Parameters | Size | Architecture | Reasoning |
|---|---|---|---|---|---|
| ChatGPT 4o | May 2024 | | | | N |
| ChatGPT 4o-mini | July 2024 | | | | N |
| Gemini 2.0 Flash | February 2025 | | | | N |
| gemma-3-27b-it | March 2025 | 27b | 17GB | Dense | N |
| gemma-3-12b-it | March 2025 | 12b | 8.1GB | Dense | N |
| gemma-3-4b-it | March 2025 | 4b | 3.3GB | Dense | N |
| gemma-3-1b-it | March 2025 | 1b | 815MB | Dense | N |
| qwen3:32b | April 2025 | 32b | 20GB | Dense | Y |
| qwen3:8b | April 2025 | 8b | 5.2GB | Dense | Y |
| deepseek-r1:32b | January 2025 | 32b | 20GB | Dense | Y |
| deepseek-r1:8b | January 2025 | 8b | 5.2GB | Dense | Y |
| Llama 4 Scout | April 2025 | 17b | 67GB | MoE (16) | Y |
| Magistral Small | June 2025 | 24b | 6GB | Dense | Y |

**ChatGPT 4o** and **ChatGPT 4o-mini** are exclusively cloud-based, and have unknown architectures, sizes and parameter counts. ChatGPT 4o-mini is a smaller and cheaper version of ChatGPT.
**Gemini 2.0 Flash** is exclusively cloud-based, with unknown architectures, sizes and parameter counts. Google's Gemini 2.0 Flash succeeded Gemini 1.5 Flash and both are cheaper and presumably smaller versions of the flagship Gemini Pro models. Gemini is multimodal for input, with text output (https://developers.googleblog.com/en/gemini-2-family-expands/).
**Gemma 3** (**gemma-3-27b-it, gemma-3-12b-it**, **gemma-3-4b-it**, **gemma-3-1b-it**) is a text/image multimodal open weights LLM from Google (https://deepmind.google/models/gemma/gemma-3/).
**Qwen 3** (**qwen3:32b**; **qwen3:8b**) is from Alibaba. The largest version has 235 billion parameters (Qwen3-235B-A22B). Alibaba claims that small variants, such as Qwen3-4B can match the performance of much larger earlier models, such as Qwen2.5-72B-Instruct (https://qwenlm.github.io/blog/qwen3/).
**DeepSeek R1** (**deepseek-r1:32b**; **deepseek-r1:8b**) is based on DeepSeek V3 with reasoning enhancements (Guo et al., 2025). DeepSeek has released smaller dense distilled open weights versions of the full model with sizes varying from 1.5B to 70B. The web version includes web search, but the open weights version does not.
**Llama 4 Scout** (llama4:16x17b) is the fourth generation meta LLM.
**Magistral Small** (magistral:24b) is the fourth generation LLM from Mistral AI, built on top of Mistral Medium 3 rather than trained natively for reasoning (Rastogi et al., 2025).

The DeepSeek models above are derived from DeepSeek V3, which was released in December 2024 and has 671 billion parameters. It uses the MoE (256 experts) architecture for efficiency, with only 37b parameters being activated on each token. Some experts are shared by all tokens and others are activated on demand. Its training dataset consisted of 15 trillion tokens (Liu et al., 2024). Its file size seems to be 685GB compressed and 1.3TB uncompressed. DeepSeek V3 seems to be approximately comparable to ChatGPT (unspecified version) in performance (Etaiwi & Alhijawi, 2025). The DeepSeek 3.1 update incorporates both reasoning and chat versions of DeepSeek V3, switching between them internally.

## *Prompting strategies*

The LLM prompts first describe the research evaluation task and then request an evaluation of a single article. The task description was taken almost verbatim from the REF instructions for assessors for Main Panel A (REF2021, 2019), except that the first sentence was changed to directly address the AI reader, "You are an academic expert" which is unnecessary for humans. This description is the Main Panel A system prompt used before (see the Appendix of: Thelwall & Yaghi, 2025) and is available in the online supplementary materials (10.6084/m9.figshare.30382651). The system prompt defines the task in terms of evaluating the originality, significance and rigour dimensions of research quality, with examples and definitions, and the scoring system of 1* to 4* mentioned above.

For models with separate user and system prompts, these instructions were submitted as the system prompt and for other models the user prompts described below were appended to them as (long) user prompts.

The main style of user prompt used was, "Score this article:\n" followed by the article title, the text "\nAbstract\n" and the article abstract. Here "\n" indicates a newline in the Json format used. Full texts were not entered because previous research has indicated that results from full text inputs give similar results, in terms of their correlations with expert scores (Thelwall, 2025bc). Using full texts would have made some of the models impractical because of the extra computational costs for processing longer inputs.

For the **few-shot** input style the user prompt was preceded by "This article scores [score]:" followed by the title, the text "\nAbstract\n", the article abstract, and the text "###\n" as a separator. The separator was added (in a format widely found in LLM outputs in the experiments) to avoid ambiguity between the examples, and to separate the main task. In a few cases the few-shot prompt confused Llama4 and Mistral and they scored all five articles (the four examples as well as the main article) but otherwise the few-shot prompt seemed to be understood (or ignored). The prompts are available in the supplementary materials (10.6084/m9.figshare.30382651).

## *Analysis*

The sole important statistic to evaluate the results is the extent to which the rank order placed on articles by the model scores aligns with the rank order of a gold standard. This is directly assessed by the Spearman rank correlation coefficient. Accuracy, precision, recall and F1 are irrelevant because the LLM scores operate on a different scale to human judgements. Note that the LLM scores produced by the above method are not integers because they average five scores and LLM scores are sometimes fractional, including when overall scores are not reported but only individual scores for rigour, originality and significance.

Bootstrapped confidence intervals were calculated for each Spearman correlation to estimate their precision. Non-overlapping confidence intervals were interpreted as evidence that one strategy is statistically significantly better than another. This is a conservative approach in the sense that a direct test of statistical significance can be passed even if there are slightly overlapping confidence intervals. Nevertheless, given the number of comparisons made, some statistically significant differences would be expected, even under this criterion, due to the large number of comparisons made (the familywise error rate issue: Benjamini, 2010).

There are six tests for each strategy, one for each of UoA 1 to 6. Given that differences between strategies are often small, a second test of statistical significance will be used: the number of UoAs for which strategy A is better than strategy B. If both are the same, then the probability that A has a higher correlation than B five out of six times is 6 choose 5, which is p=0.09. In contrast the probability for 6 out of 6 is 0.016, so 6 is taken as statistical evidence that strategy A is better than strategy B. Similarly, at least 7 out of 8 successes (p=0.031) are needed to pass the p=0.05 threshold for statistical significance for the cases where eight models are analysed with two different approaches, and 9 out of 10 (p=0.011) for RQ3, and 13 out of 16 for RQ4 (p=0.0176).

When making statistical or AI estimates it is common to analyse the errors with a distribution plot or confusion matrix. Neither is appropriate here for the individual article scores because one variable is (almost) ordinal and the other is scalar. Thus, violin plots were used instead to show the range of predicted scores for each individual score, discarding the few non-integer scores in the second gold standard. In the violin plots, the median is shown with a black line, the range is shown with whiskers, and the density of the results is indicated by the violin shape, which is a smoothed estimate from the data.

Violin plots were not drawn for the departmental average data because it is less informative since the x axis would not include any claimed correct scores, just proxies. There is also less data for each x axis value, with none being integers.

An additional procedure was conducted to give insights into model differences, reasoning processes, and the effect of few-shot: Word Association Thematic Analysis (WATA) (Thelwall, 2021). This is a hybrid qualitative-quantitative method to identify themes in the common differences between two sets of texts. It is a statistically robust method because it only investigates words that have statistically significant frequency differences between the two texts (with a chi-square test, backed by a Benjamini Hochberg procedure to protect the familywise error rate: Benjamini & Hochberg, 2000). The qualitative stage is to cluster together words with similar contexts into themes. These contexts are identified by reading a random sample of texts and the themes are identified by clustering related terms, based on their contexts, with an inductive thematic approach. For this article a light version was used, identifying only the main themes rather than all themes because there were too many comparisons.

WATA was used to compare the reports produced by the different medium sized models against each other in pairs, and to compare few-shot and zero-shot reports for the same model. For each comparison, all five iterations were combined and since two sets were compared against each other, so each WATA dataset contained 5 x 4780 x 2 = 47,800 model reports. To avoid reporting too many different analyses, the reasoning models were compared against each other, the non-reasoning models were compared against each other and only one non-reasoning model, Gemma 3 27b, was compared against all the reasoning models to identify the main type differences.

Finally, to check whether a weighted sum of the different models scores could produce better results than an individual model, evolutionary optimization with differential evolution (Lampinen & Storn, 2004) was used to identify sets of non-negative weights to optimize the Spearman correlation with the gold standard, using 10-fold cross-validation to guard against overfitting. This was a supplementary test with the available data. There is insufficient data for a holdout set, so, to guard against overfitting, only the simplest option was used (e.g., including all models, using only one fitting mechanism).

# Results

## *RQ1 large vs medium*

All medium sized LLMs give similar correlations to the cloud-based LLMs and, based on the confidence interval estimates, there is insufficient evidence to conclude that cloud-based LLMs are superior for this task overall (Figure 1). Moreover, the reasoning models Magistral Small, Qwen3 32b, DeepSeek R1 32b perform well but do not have a clear advantage over the similar-sized non-reasoning Llama4 Scout and Gemma3 27b. The conclusion is not affected by the choice of gold standard. The overall pattern is similar within UoAs 1 to 6 (Figures 2 and 3).

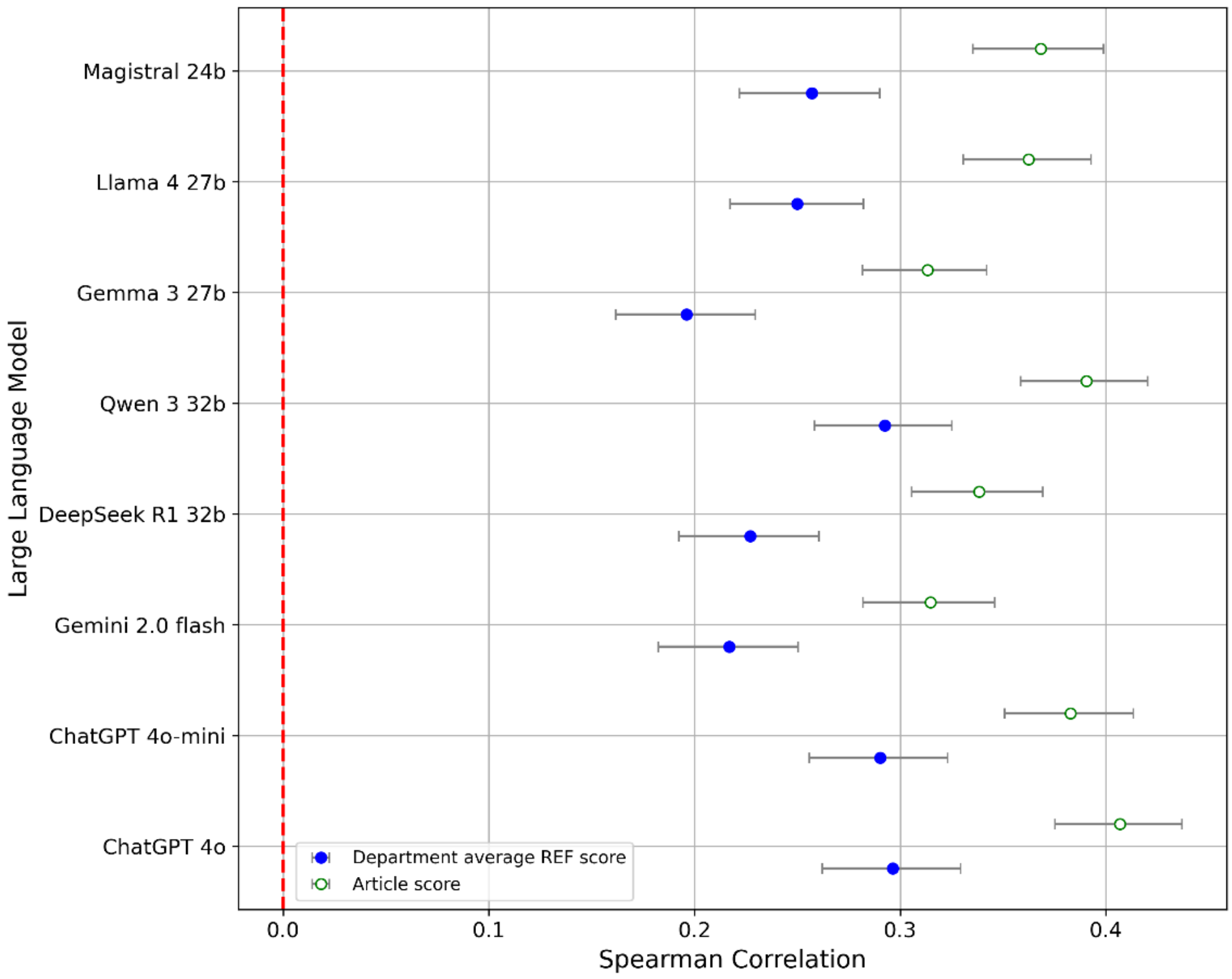

Figure 1. Average Spearman correlations between departmental average REF scores or article score estimates and LLM scores for a selection of large and medium sized LLMs. The correlations are averaged over 500 articles in each of UoAs 1 to 6, and the error bars are 95% confidence interval estimates from these.

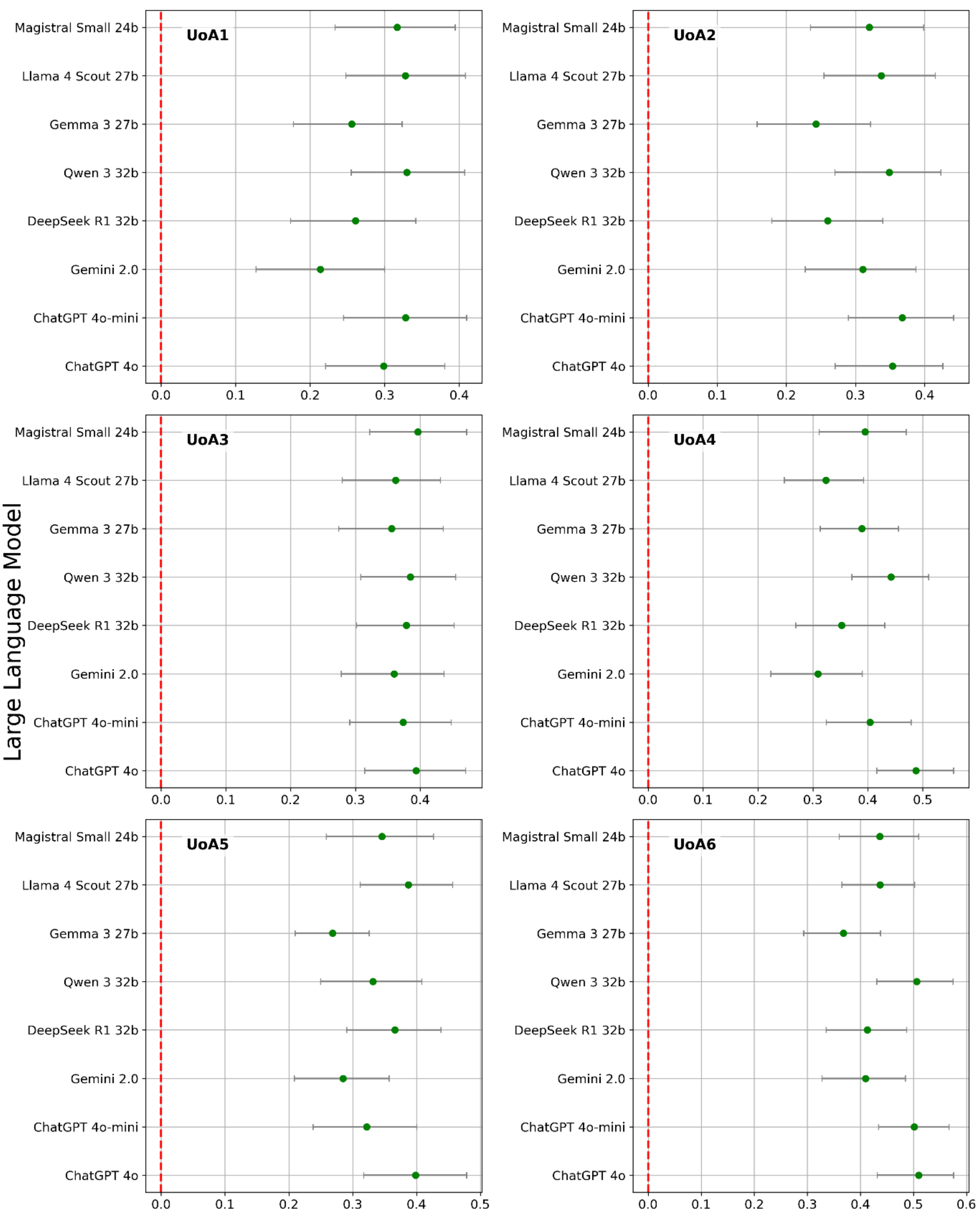

Figure 2: Average Spearman correlations between article score estimates and LLM scores for a selection of large and medium sized LLMs. The error bars are bootstrapped 95% confidence intervals. The UoAs are 1 Clinical Medicine; 2 Public Health, Health Services and Primary Care; 3 Allied Health Professions, Dentistry, Nursing and Pharmacy; 4 Psychology, Psychiatry and Neuroscience; 5 Biological Sciences; 6 Agriculture, Food and Veterinary Sciences.

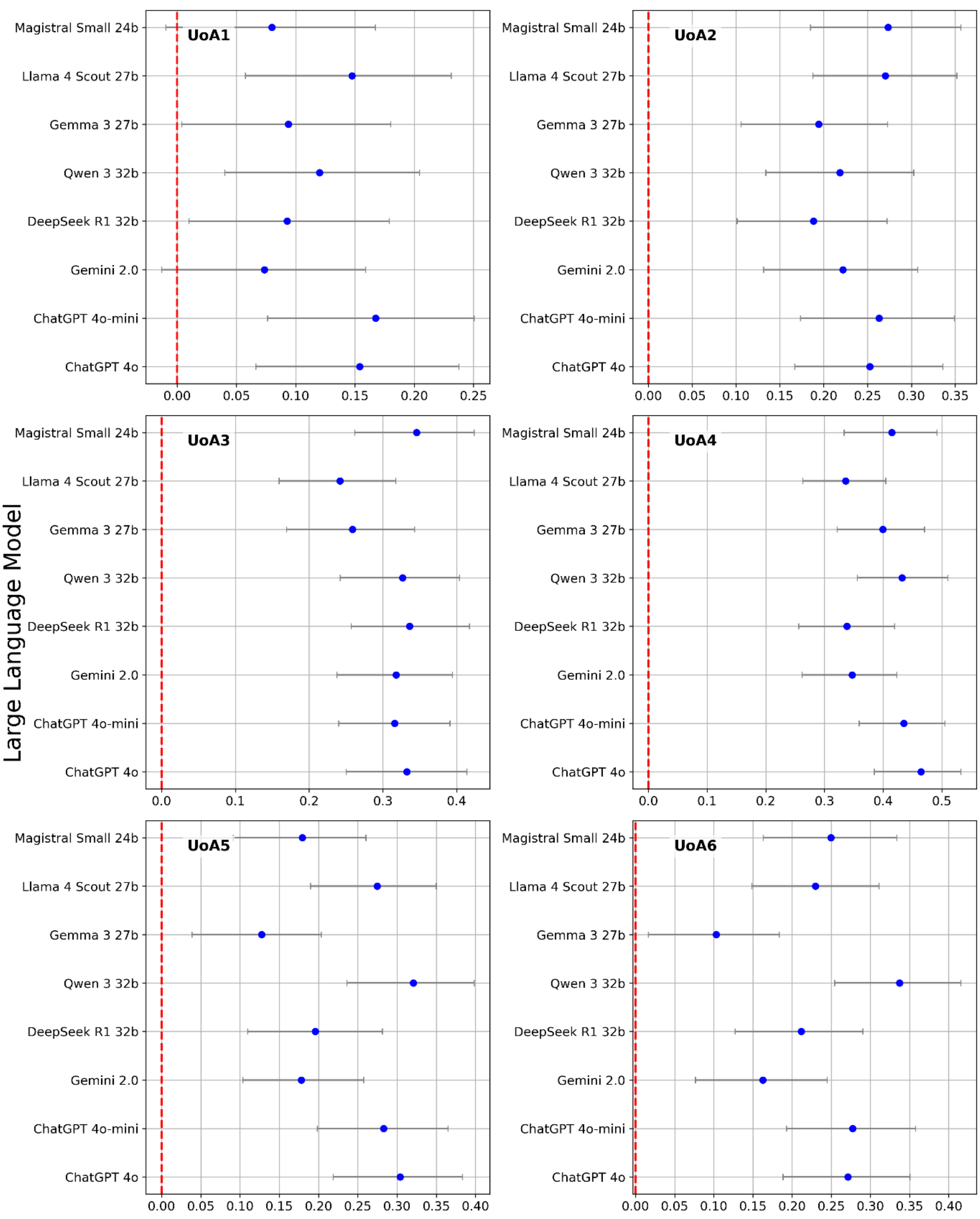

Figure 3: Average Spearman correlations between departmental average REF scores and LLM scores for a selection of large and medium sized LLMs. The error bars are bootstrapped 95% confidence intervals. The UoAs are 1 Clinical Medicine; 2 Public Health, Health Services and Primary Care; 3 Allied Health Professions, Dentistry, Nursing and Pharmacy; 4 Psychology, Psychiatry and Neuroscience; 5 Biological Sciences; 6 Agriculture, Food and Veterinary Sciences.

## *RQ2 medium vs small*

Choosing a smaller LLM in a family tends to weaken the correlations but not substantially, except for Gemma 3 1b. Thus, the minimum practical size for an LLM may be between 4b and 1b, although the cutoff may be different for LLM families other than Gemma 3 (Figure 4). The conclusion is not affected by the choice of gold standard. The patterns within UoAs 1 to 6 are broadly consistent with the overall pattern, except that within UoAs 2 and 5, low correlations suggest that 4b might be too small (Figures 5 and 6).

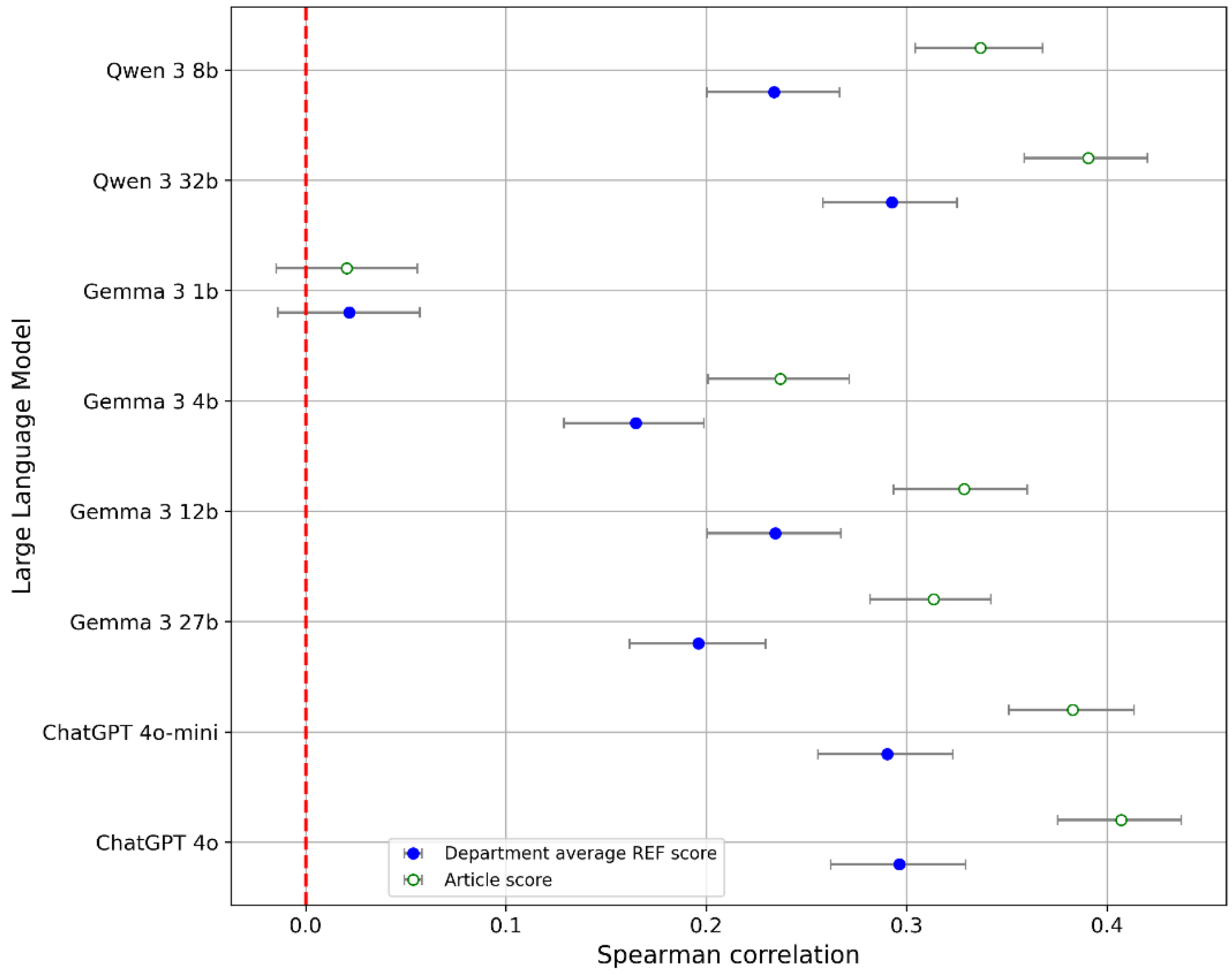

Figure 4. Average Spearman correlations between departmental average REF scores or article score estimates and LLM scores for papers with a selection medium sized and smaller variant LLMs. The correlations are averaged over 500 articles in each of UoAs 1 to 6, and the error bars are 95% confidence interval estimates from these.

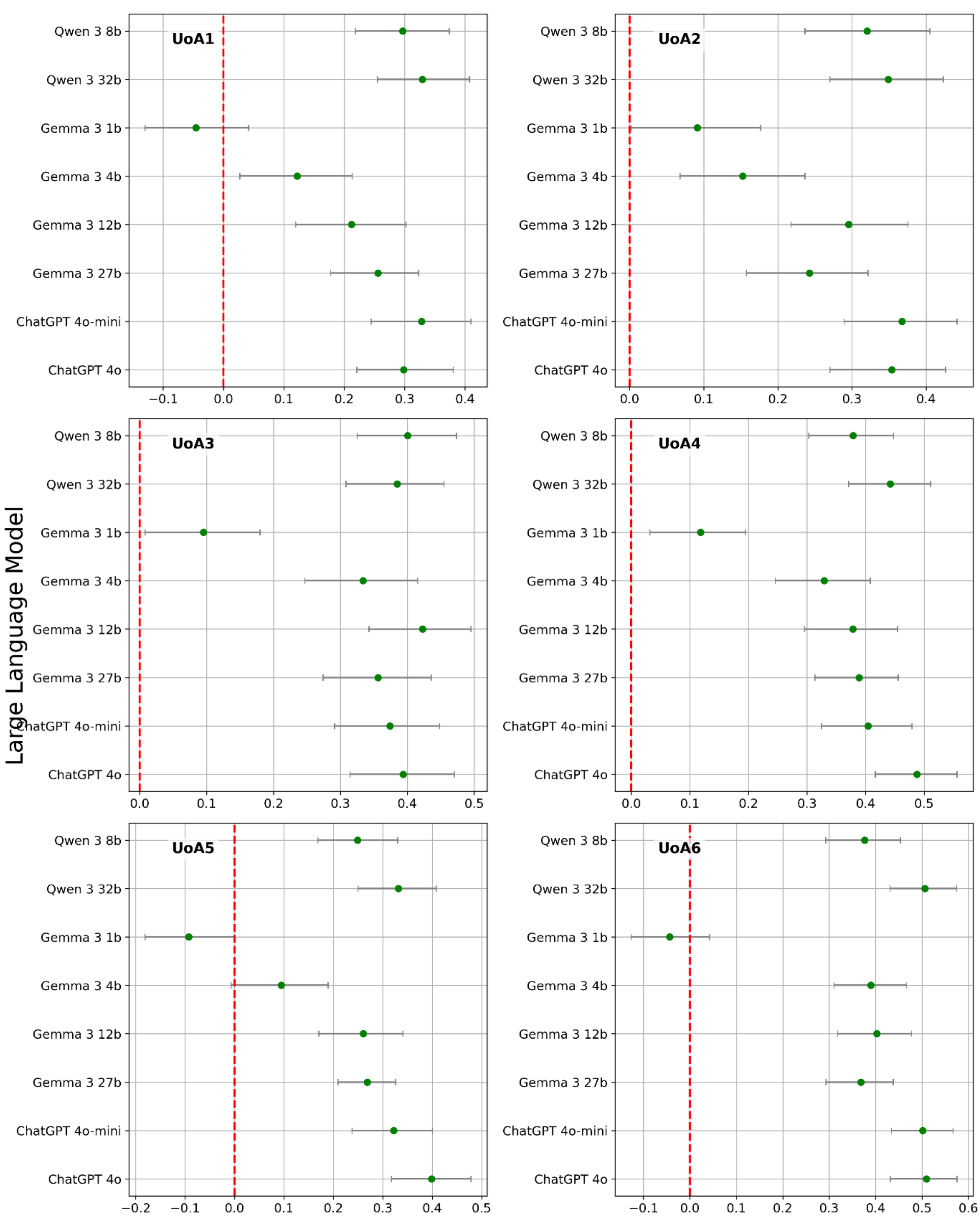

Figure 5. Average Spearman correlations between article score estimates and LLM scores for papers with a selection medium sized and smaller variant LLMs. The error bars are bootstrapped 95% confidence interval estimates. The UoAs are 1 Clinical Medicine; 2 Public Health, Health Services and Primary Care; 3 Allied Health Professions, Dentistry, Nursing and Pharmacy; 4 Psychology, Psychiatry and Neuroscience; 5 Biological Sciences; 6 Agriculture, Food and Veterinary Sciences.

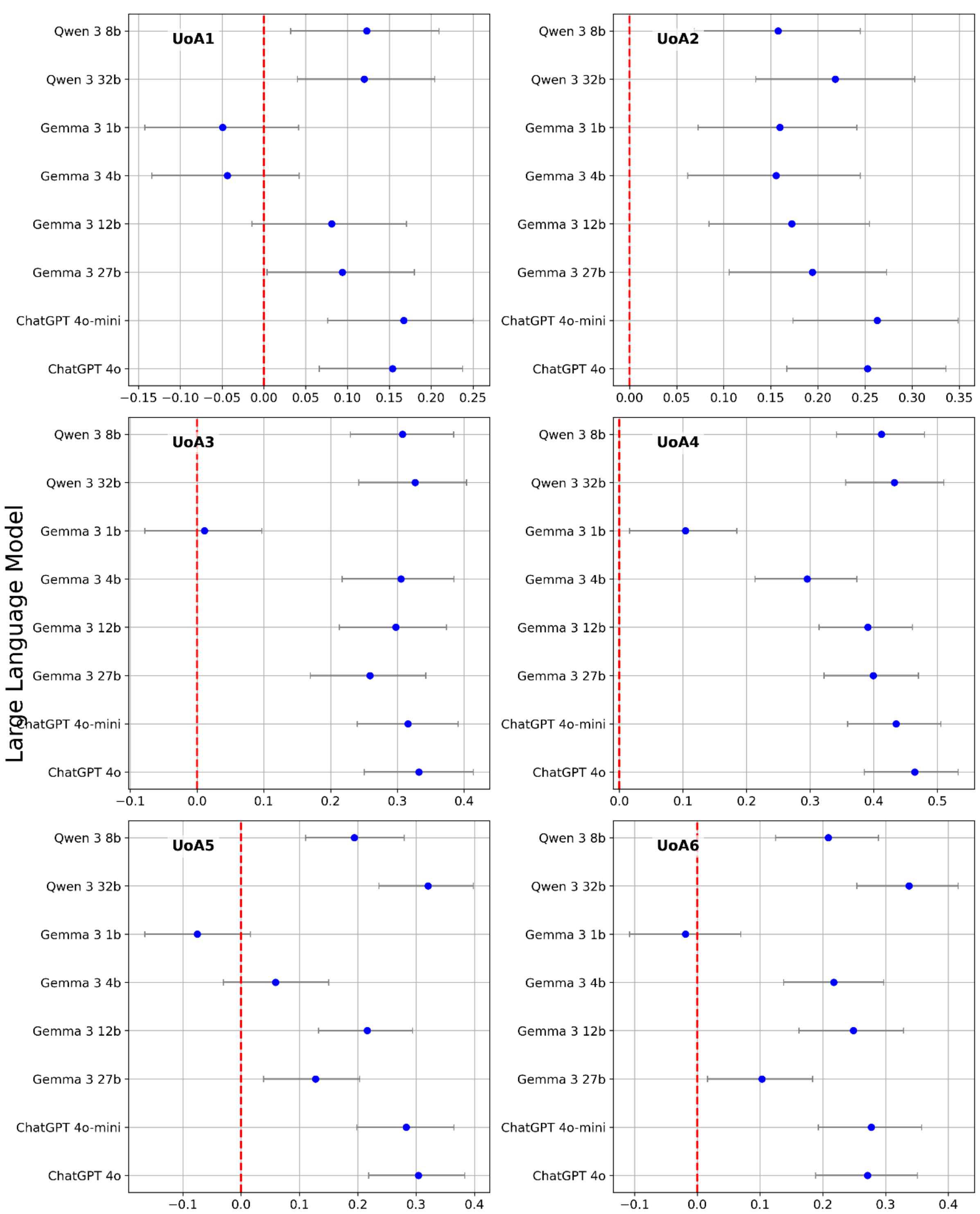

Figure 6. Average Spearman correlations between departmental average REF scores and LLM scores for a selection medium sized and smaller variant LLMs. The error bars are bootstrapped 95% confidence interval estimates. The UoAs are 1 Clinical Medicine; 2 Public Health, Health Services and Primary Care; 3 Allied Health Professions, Dentistry, Nursing and Pharmacy; 4 Psychology, Psychiatry and Neuroscience; 5 Biological Sciences; 6 Agriculture, Food and Veterinary Sciences.

## *RQ3 zero-shot vs few-shot*

The few-shot strategy tried here increased the correlation in 6 out of 10 cases assessed (9 out of 10 for the departmental average score data), and sometimes substantially (Gemma 3 27b, Gemini 2.0 flash, DeepSeek R1 32b). Thus, whilst the evidence is weak, it tends to support the few-shot strategy. The poor results for Llama 4 may be due to it sometimes misinterpreting the examples and classifying all five titles/abstracts instead of just the target. The same occurred occasionally for Mistral. Thus, some LLMs may be less able to take advantage of the few-shot method. The conclusion is not affected by the choice of gold standard. The patterns within UoAs 1 to 6 are broadly consistent with the overall pattern (Figures 8 and 9).

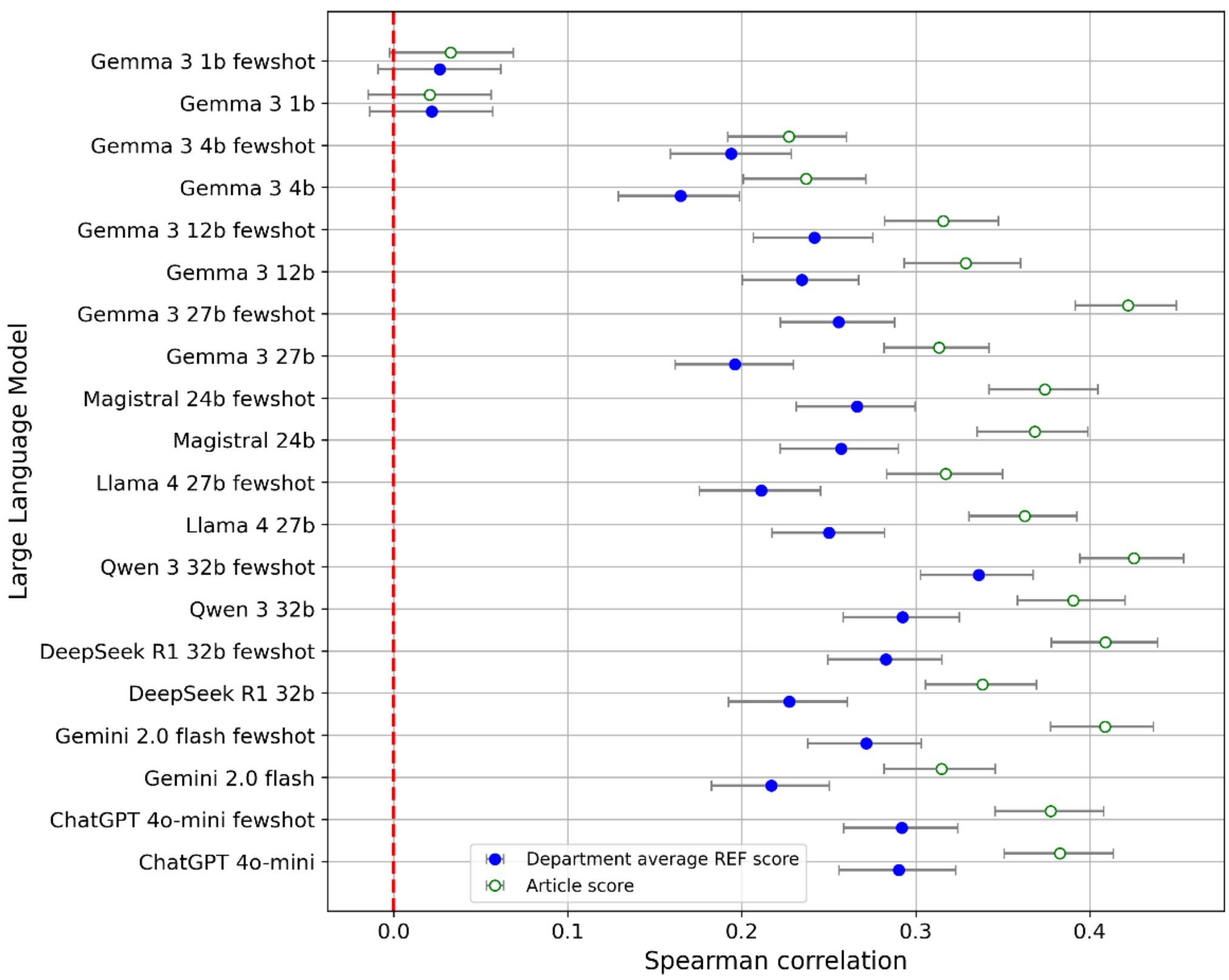

Figure 7. Average Spearman correlations between departmental average REF scores or article score estimates and LLM scores for papers with a selection of larger and medium sized LLMs with zero-shot and few-shot prompting. The correlations are averaged over 500 articles in each of UoAs 1 to 6, and the error bars are 95% confidence interval estimates from these.

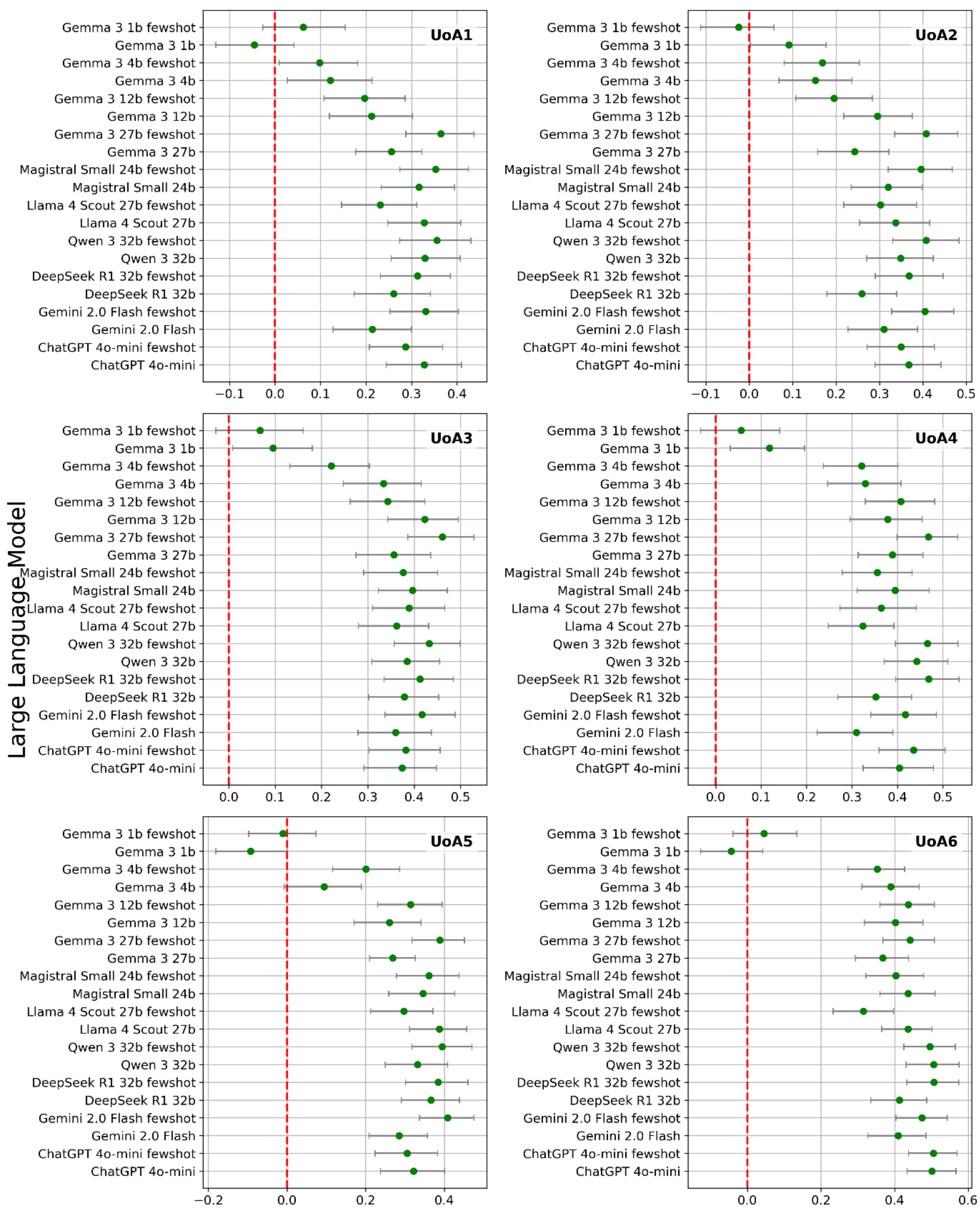

Figure 8. Average Spearman correlations between article score estimates and LLM scores for papers with a selection of larger and medium sized LLMs with zero-shot and few-shot prompting. The error bars are bootstrapped 95% confidence intervals. The UoAs are 1 Clinical Medicine; 2 Public Health, Health Services and Primary Care; 3 Allied Health Professions, Dentistry, Nursing and Pharmacy; 4 Psychology, Psychiatry and Neuroscience; 5 Biological Sciences; 6 Agriculture, Food and Veterinary Sciences.

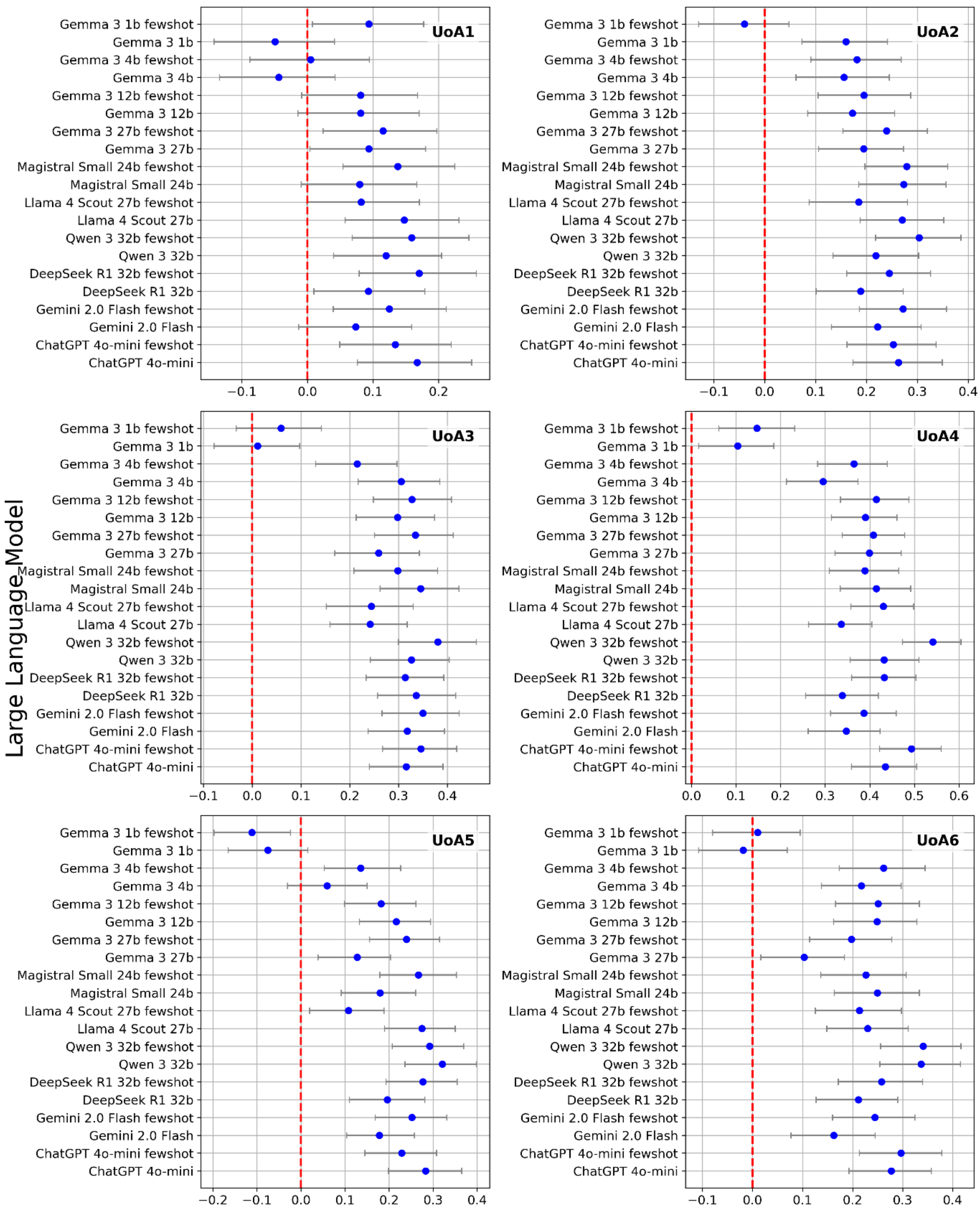

Spearman Correlation against departmental average scores

Figure 9. Average Spearman correlations between departmental average REF scores and LLM scores for a selection of larger and medium sized LLMs with zero-shot and few-shot prompting. The error bars are bootstrapped 95% confidence intervals. The UoAs are 1 Clinical Medicine; 2 Public Health, Health Services and Primary Care; 3 Allied Health Professions, Dentistry, Nursing and Pharmacy; 4 Psychology, Psychiatry and Neuroscience; 5 Biological Sciences; 6 Agriculture, Food and Veterinary Sciences.

### *RQ4 average vs individual*

In all 16 contexts tested, averaging five LLM scores gives a higher Spearman correlation than taking a single score (Figure 10). Thus, the averaging strategy used throughout this article seems to be widely effective. This conclusion is not affected by the choice of gold standard.

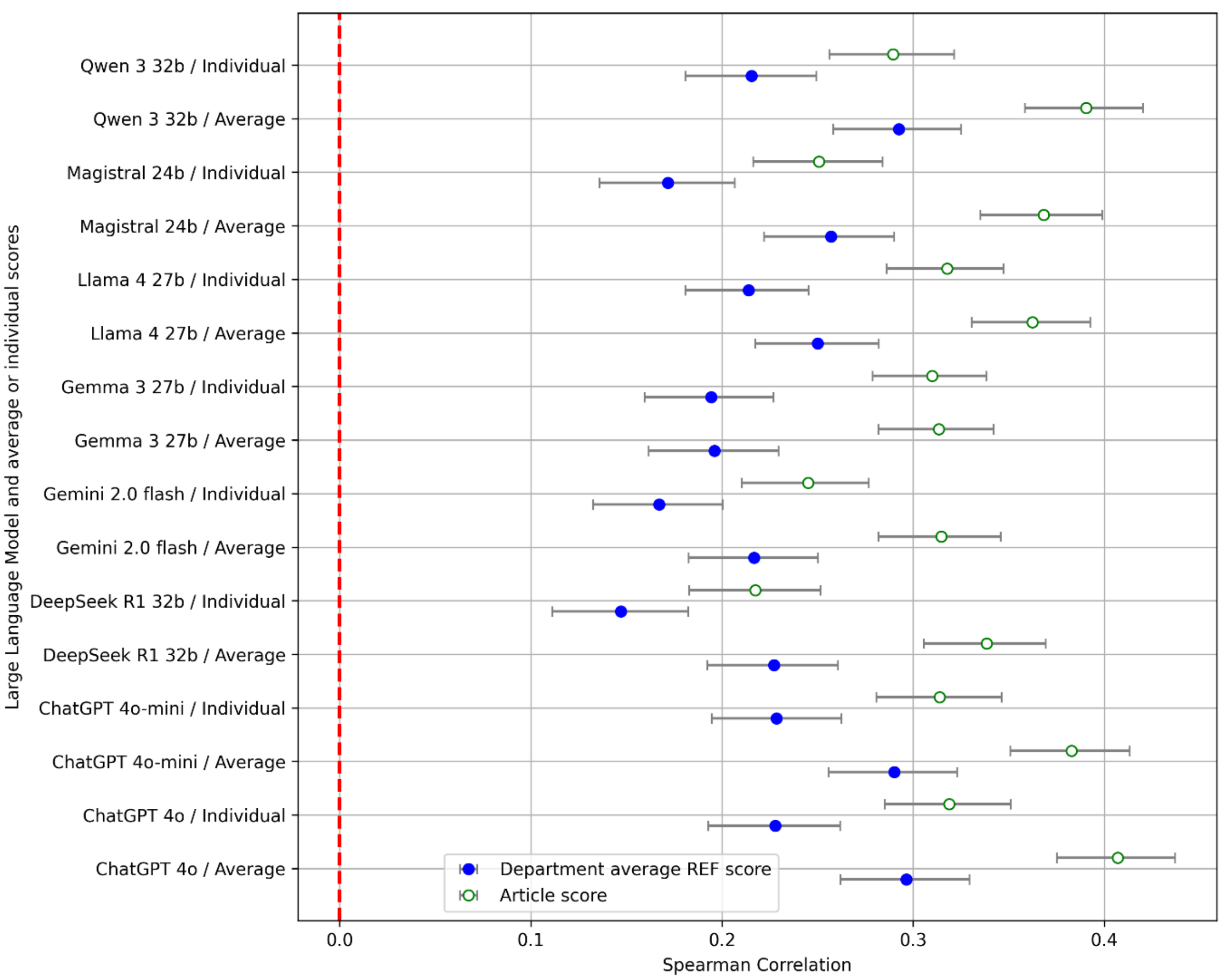

Figure 10. Average Spearman correlations between departmental average REF scores or article score estimates and LLM scores for papers with a selection of larger and medium sized LLMs individual LLM score estimates or estimates based on an average of five scores per model. The correlations are averaged over 500 articles in each of UoAs 1 to 6, and the error bars are 95% confidence interval estimates from these.

## Discussion

The results are limited in multiple dimensions. The dataset is UK-focused and both gold standards are imperfect: one is indirect, and the other is subjective to a single individual. It also covers only the health and life sciences, although in the UK context these accounted for 41% of journal articles in REF2021. Of course, there are certainly disciplinary differences in the value of LLMs for research assessment. Perhaps most importantly (and frustratingly), (a) the confidence intervals are relatively wide compared to the differences found, despite the sample sizes of 500, so the differences may be artefacts of the datasets used, and (b) LLMs have different architectures even within broad parameters (e.g., “reasoning model”) and so even statistically significant differences for one model may not transfer to another ostensibly similar one. Also, only a single performance metric has been reported, Spearman correlation, whereas LLM studies usually include a range to give wider performance information. In the current case, since the individual scores are inaccurate, only the rank correlation is important.

Another limitation is that the LLMs compared became available years after the REF2021 articles were published and this may influence their judgements of novelty, because a novel contribution in 2014 would not be novel in 2023. Thus, the correlations reported might be higher if applied to contemporary research because the novelty evaluations would be more effective. Nevertheless, age has only a small effect on the scores overall (Thelwall & Kurt, 2025), so this probably has not reduced the correlations reported here substantially.

In terms of model differences, the results do not give strong evidence of the capabilities of LLMs that were not included because their different architectures may yield divergent results. Moreover, future LLMs may work differently enough that new patterns emerge in the results.

For efficiency, all the Gemma3 results in the current paper were obtained from API calls to Google rather than from local copies. As a sanity check, the Gemma 3 27b results were also run locally, and the correlation was higher than reported above (average correlation), but with overlapping confidence intervals. This would change the results but is not reported since this was not part of the original research design. Thus, the Gemma 3 27b main correlations may underestimate its performance relative to the other models (including the Gemma 12b, which outperformed it. This extra information supports the commonsense conclusion that the correlations tend to be universally weaker with smaller model variants from the same family.

## *Model size (RQ1, RQ2)*

Although it is known for other tasks that larger LLMs tend to be more effective (Sardana et al., 2023) and that there can be emergent properties that effectively don't exist if the model is too small (Berti et al., 2025), no previous study has investigated the relationship between LLM size and effectiveness at the task of academic research quality evaluation.

The results confirm the previous finding, with the single model Gemma 3 27b, that the largest (typically cloud-based) LLMs are not necessary for this task (Thelwall, 2025a). The results go further by showing that other open weights models have the same property and that even smaller models, such as Gemma 3 12b can work reasonably well and even the smaller Gemma 3 4b for some fields. The results also suggest that Gemma 3 1b is too small. Since no other tests of similar size models were conducted to assess the extent to which the minimum size is model dependant, it remains speculation that a minimum size of between 4b and 12b is generally necessary to work for all fields.

## *Few-shot (RQ3)*

Previous research on other problems has found few-shot to work well (Wu et al., 2025) although there are some concerns about the validity of the conclusions (Li & Flanigan, 2024). The current experimental setup seems immune to data leakage from few-shot because the few-shot samples were not part of the gold standards.

Although few-shot did not always improve the correlations, the weak evidence that it tended to improve the results more often than not suggests that this approach should be investigated further, such as with different strategies (e.g., number of articles; the way of presenting the results). For example, there might be a strategy that is less likely to confuse some of the LLMs.

One way in which few-shot might be expected to influence the results is by increasing the likelihood that low scores were given, since each few-shot set included an example of a 1* article (there were no 1* scores given by the LLMs). The same is true for 2* articles since this score was also rare. Nevertheless, the average few-shot score did not tend to be lower than for the zero-shot prompt (Figure 11) and so this did not happen systematically.

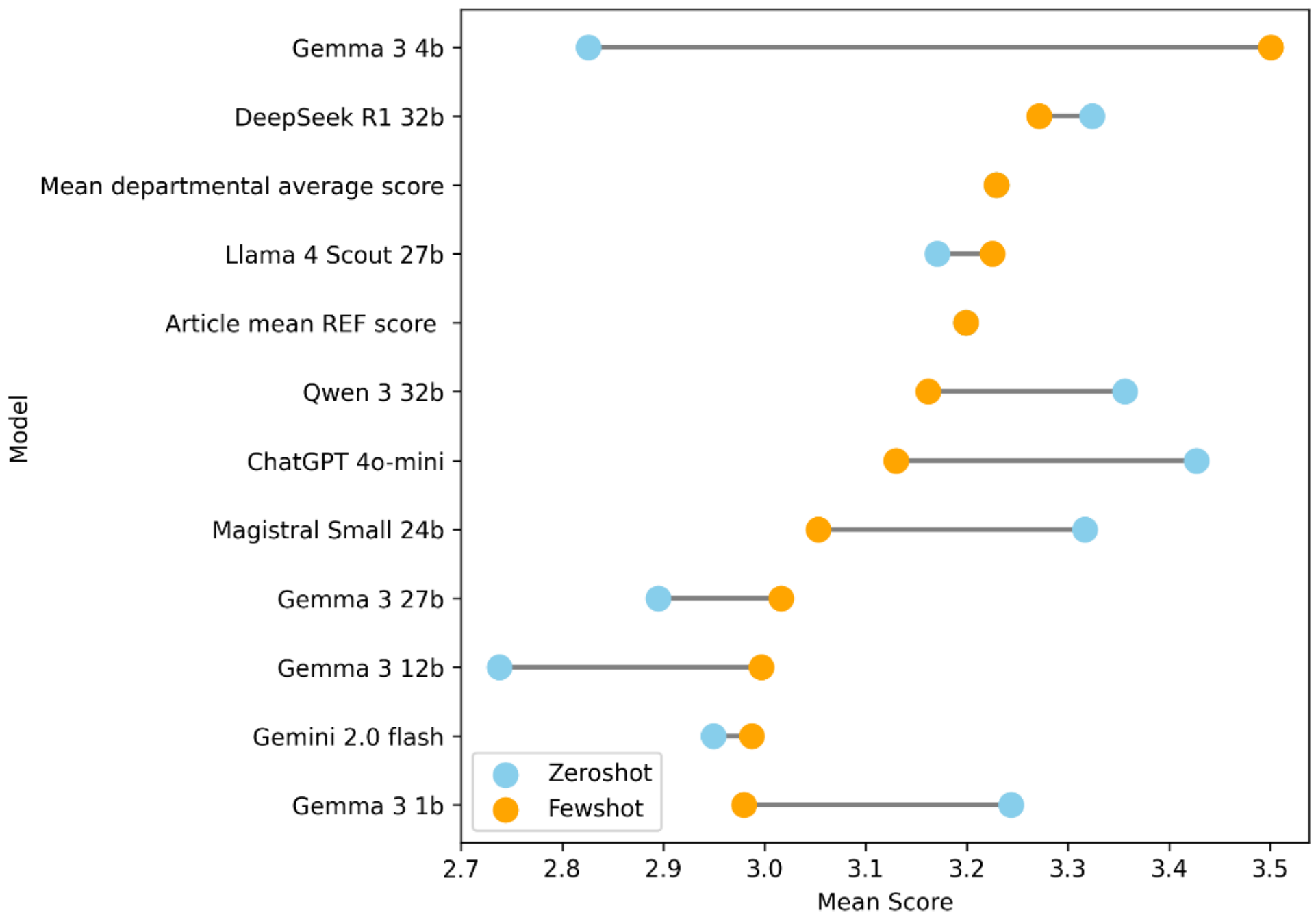

Figure 11. Mean scores for zero-shot and few-shot from a range of models and the two gold standards.

The few-shot prompt tended to elicit a wider range of scores for each article. This can be seen to a weak extent for Qwen 3 (Figure 12 against Figure 13) and to a great extent for Llama 4 (Figure 14 against Figure 15; violin plots for other models are in the supplementary materials: 10.6084/m9.figshare.30382651). A possible explanation for the suggested improved performance for the few-shot method is therefore that the strategy used altered the prompts for each of the five iterations averaged and this additional prompt variety helped generate a wider range of scores for the same article. This may be a more powerful factor than any learning from the four article scores in each case. If the few-shot examples did not help the LLMs then this suggests that a technique of effectively adding noise into a prompt may help to increase the value of averaging scores from multiple LLM prompt iterations through increased score variety.

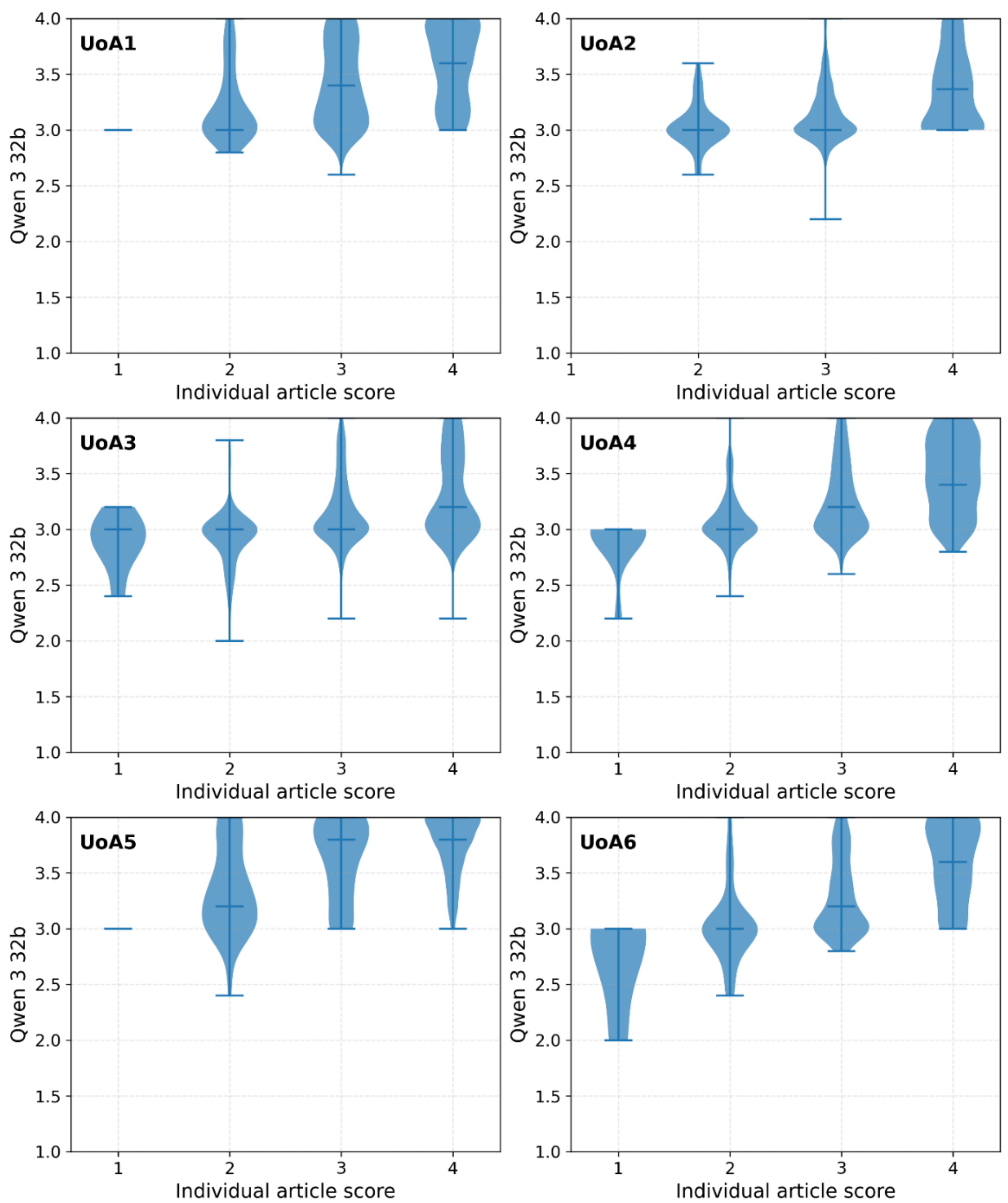

Figure 12. Violin plots of Qwen 32b score estimates (average of 5) against individual article scores for papers from UoAs 1 to 6, excluding articles with fractional scores.

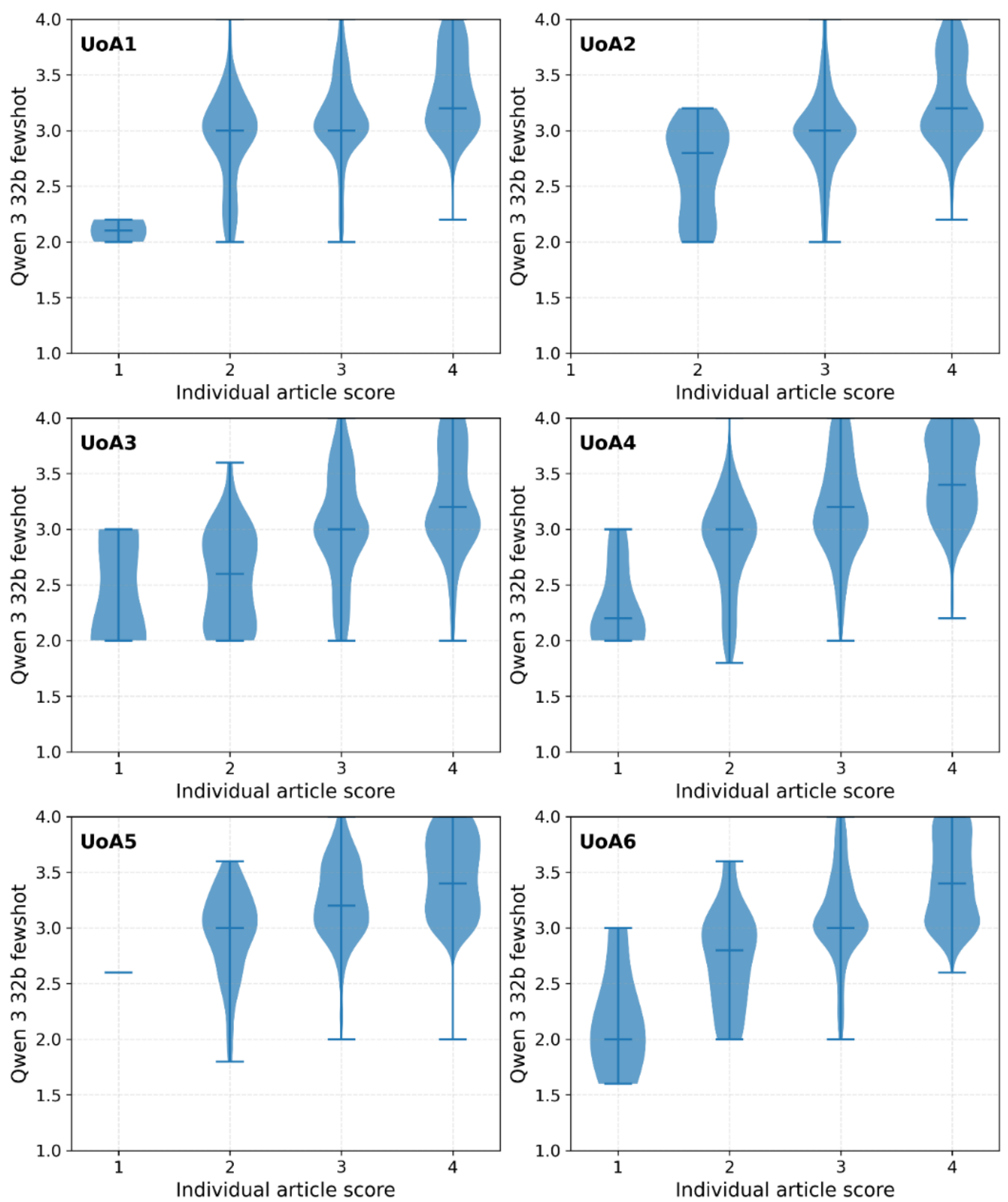

Figure 13. Violin plots of Qwen 32b few-shot score estimates (average of 5) against individual article scores for papers from UoAs 1 to 6, excluding articles with fractional scores.

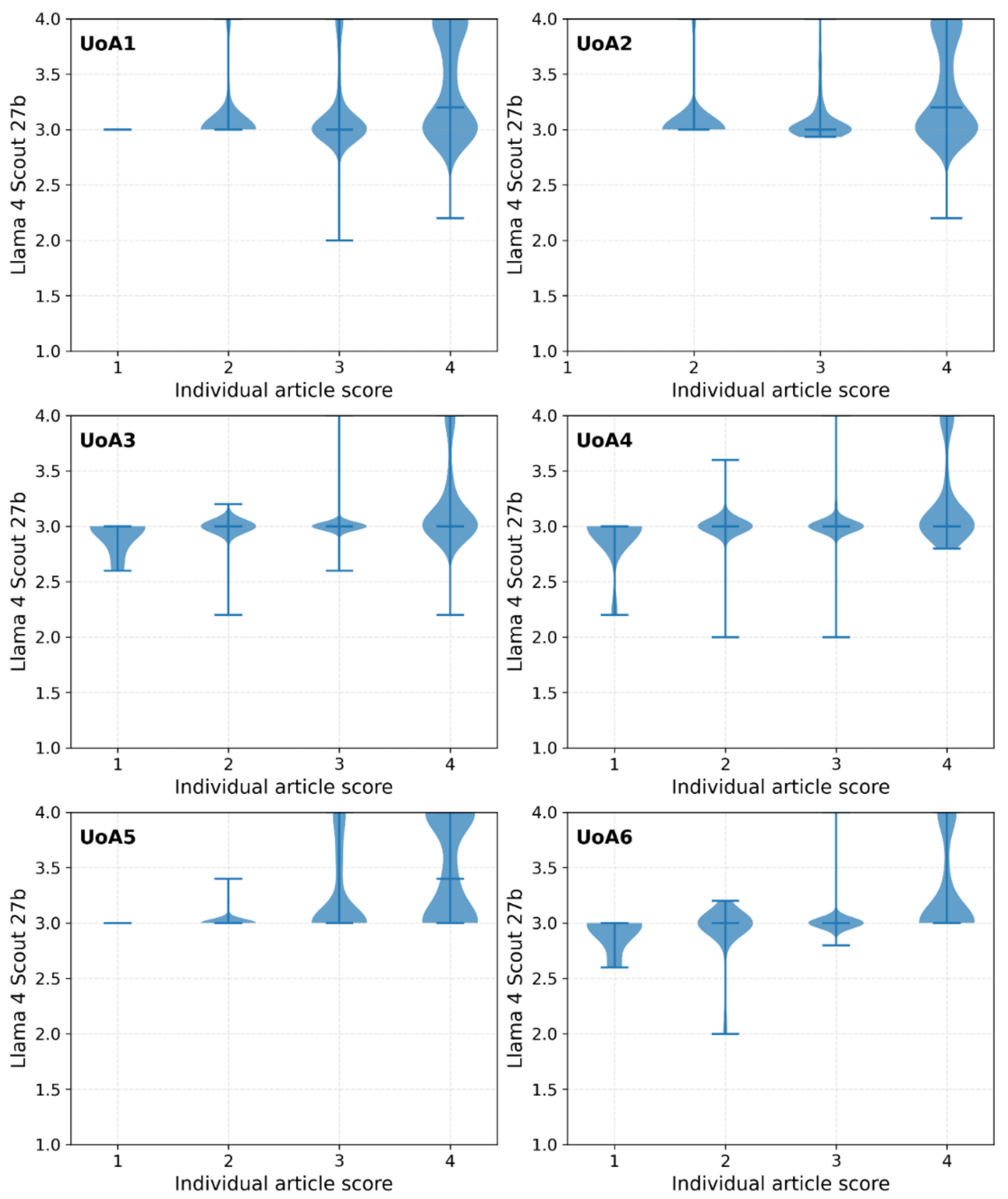

Figure 14. Violin plots of Llama 4 Scout score estimates (average of 5) against individual article scores for papers from UoAs 1 to 6, excluding articles with fractional scores.

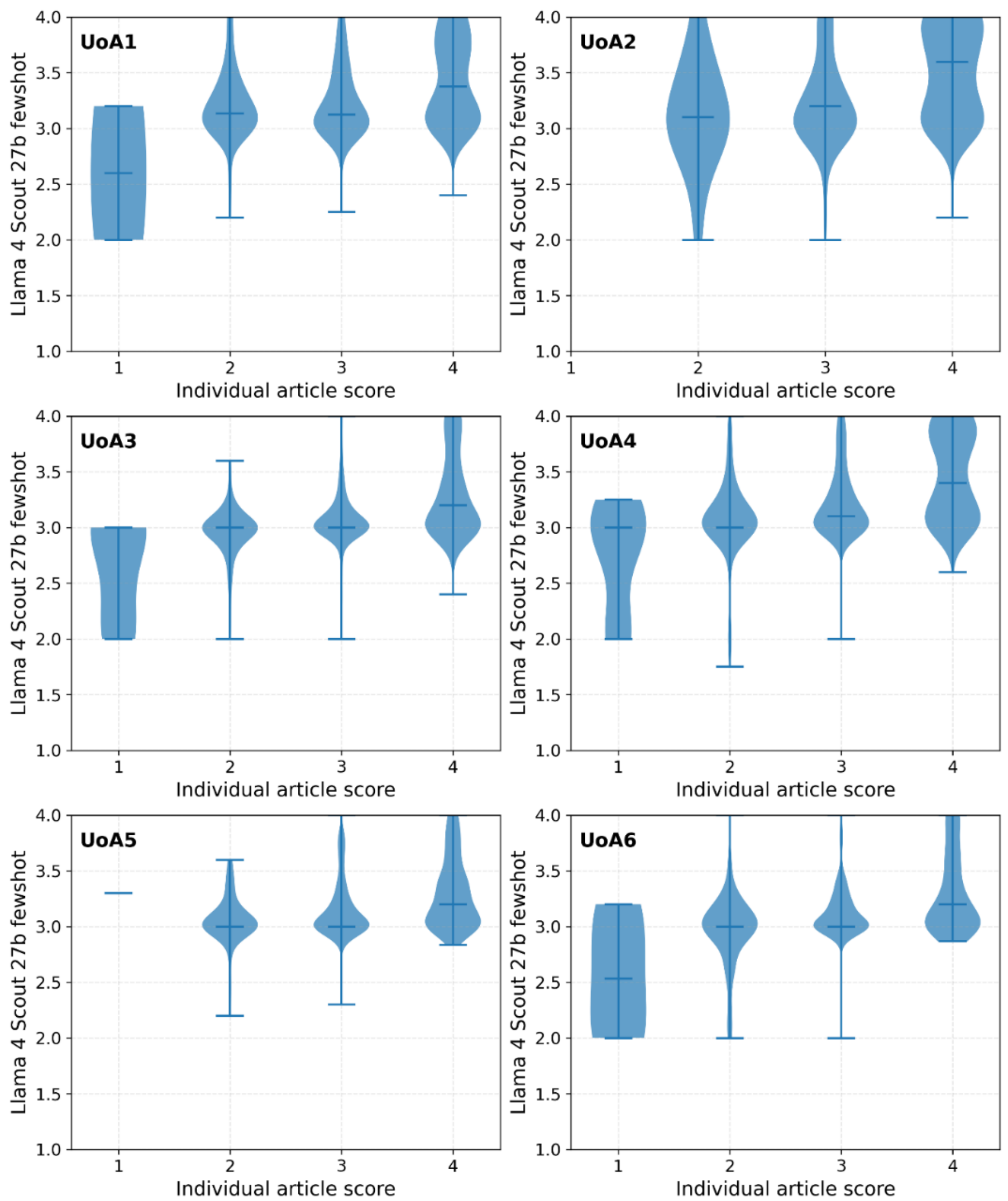

Figure 15. Violin plots of Llama 4 Scout few-shot score estimates (average of 5) against individual article scores for papers from UoAs 1 to 6, excluding articles with fractional scores.

## *Averaging (RQ4)*

Averaging scores from multiple identical LLM prompts is a rare technique for LLMs in general but has been previously shown to be effective for this task (Thelwall, 2024) and subsequent studies have assumed this without testing. The current paper gives robust evidence that this assumption has general validity for different models. This is not surprising and aligns with a previous hypothesis that averaging multiple scores from identical prompts is a way to leverage model internal probability information that is less fully used if only a single score is taken (Thelwall & Yang, 2025).

## *Reasoning*

Although not explicitly addressed in a research question, the results give a chance to compare new reasoning and non-reasoning models. They had similar performances for this task and had similar sizes, so it seems that reasoning is not necessary for research quality scoring from titles and abstracts. Not all tasks require reasoning and even though research quality scoring is an expert task with complex rules, the relatively sparse information in abstracts may result in a more intuitive/pattern matching approach working as well as reasoning. In other words, even though the instructions are complex, there may be too little to reason about. In contrast, reasoning models seem likely to work well on multi-step problems, such as mathematical and logical puzzles (Jahin et al., 2025), coding, and reasoning with supporting documents (at least for o1: Wu et al., 2024). They may also help text generation with a set of compositional constraints (Wu et al., 2024), for which checking steps may help. Although such additional checking steps could help for research assessments, such as for norm referencing or benchmarking (see below) since “self-reflection” is a strength (Plaat et al., 2025), the sparseness of the information to check and the difficulty of making norm referencing type comparisons for diverse research texts, seems to make this ineffective.

One side-effect of reasoning was a greater spread of scores, as suggested by comparisons of the violin plots of Qwen 3 32b and Llama 4 Scout above, and similar comparisons of violin plots in the supplementary materials (10.6084/m9.figshare.30382651). Reasoning models were also approximately 5-10 times slower, which is a substantial practical consideration mitigating against their use in the absence of a clear advantage.

Qwen 3 and DeepSeek R1 32b report reasoning as part of their output. The discussion in the reasoning section seems to primarily reflect factors that are usually included in the non-reasoning report, rather than drawing upon a new strategy (see Appendix A). Some of the reasoning sections were very long, seeming to repeatedly re-examine the core task without making any progress. Thus, there did not seem to be a clear advantage for reasoning, at least for the standard prompts.

### *Reasoning and few-shot*

The Qwen3 32b and DeepSeek R1 32b reasoning output for few-shot prompts gives the opportunity to investigate how few-shot examples are interpreted. They were frequently explicitly used for benchmarking in the thinking sections of Qwen3, which is correct (e.g., Appendix B). The reasoning model without an explicit reasoning section in its reports, Magistral, sometimes mentioned the few-shot benchmark articles but the non-thinking model Gemma3 did not seem to ever mention the benchmarking. The non-reasoning Llama 4 Scout did occasionally mention the examples, but sometimes incorrectly interpreting them.

Llama 4 Scout misinterpretations could involve incorrect benchmark scores (e.g., “In comparison to the other articles, this article would score: * **Originality: 3/4** (Article 1: 2/4, Article 2: 2.5/4, Article 3: 3.5/4)”: no few-shot benchmark scores were fractions), but usually entailed scoring the example articles as well as the main one (e.g., “Based on the provided abstracts, I will evaluate each article's originality, significance, and rigour, and then provide a score out of 4.”) so it seemed to struggle with understanding the few-shot prompt. DeepSeek R1 32b also misinterpreted the prompt at least once (e.g., “<think> Alright, so I've been given four academic journal articles to score based on originality, significance, and rigour using a 1* to 4* scale. []</think>[]**Final Scores:** - Article 1: 2* - Article 2: 3* - Article 3: 3* - Article 4: 4*”).

Here are some comparison-focused extracts from Qwen3 32b’s thinking <think>...</think> sections:

- This study surpasses the examples provided in the 3* and 2* categories by offering a **novel, mechanistic insight** with direct therapeutic implications. Unlike the 3* study (tDCS in CP), which validates an intervention with moderate novelty, or the 2* study (sperm clustering), which explores a biological phenomenon, this work advances **theoretical frameworks** in oncology and proposes a **translatable therapeutic strategy**. The 4* classification is justified by its alignment with world-leading standards in originality, significance, and rigour.

- Comparing to the 4* example, the hydrogen peroxide article proposed a new abiotic pathway, which is a major theoretical contribution. This article is more about empirical validation of existing theories in a new context. So maybe it's 3* or 2*? The 3* example on planarians had clear findings but maybe less applicability. The 2* example improved a tool. The user's example for 3* was about regional differences in planarians with genomic instability. It provided new insights into mechanisms but maybe not as broadly applicable. This article's findings are more general and could have significant implications for forest management and policy.
- Comparing to the examples, the 4* article had a large cohort and longitudinal data, which is different. The 3* article in neuroimaging had significant findings but maybe less impact. The 2* and 1* articles were more applied or had smaller scopes.
- Wait, the user's examples show that the 4* article had a systematic review with a large number of studies and a global impact. This study is strong but perhaps not at that level. So, I'll assign it a 3*.

As the above examples illustrate, the variety of research topics even within a single UoA makes few-shot comparisons potentially unhelpful. This may explain why few-shot was not always clearly advantageous. Of course, the comparisons may help sometimes, as the following example from DeepSeek 32b illustrates, “<think>But looking back, in the example provided earlier, a similar study scored 3* overall. Considering this, maybe the overall score should be 3*. But wait, another perspective: The significance is high because it resolves uncertainty from smaller trials, so it's internationally excellent but not world-leading since the outcome was negative and might not have as broad an impact as initially hoped. So perhaps a 3* for all criteria.[]</think>”.

### *Comparison of model reports with and without few-shot*

For all the medium sized models, the few-shot reports were systematically compared with the zero-shot reports for the same model using the WATA method. The following themes were identified as more prevalent in few-shot reports compared to zero-shot.

- **DeepSeek R1 32b**: *Recapping the problem* (e.g., “The user has provided examples with scores from 1* to 4*, so I should follow that structure.”). Comparison with examples (e.g., “Comparing this to the examples: The 4* example discovered a novel genetic cause of a disorder, while this study is more about mapping and risk factors.”). *Style differences*, such as explicitly stating the title of the article evaluated.
- **Qwen3 32b**: *Recapping the problem* (e.g., “The user provided examples of articles with scores 1 to 4, so I need to align this article with those examples.”). *Comparison with examples* (e.g., Comparing to the previous examples: The 4* article introduced a new endophenotype in a genetic study.”).
- **Gemma 3 32b**: Only *style differences*, such as starting a report with “## Evaluation of:”.
- **Llama4 Scout**: *Comparison with examples* (e.g., “In comparison to the other articles provided, this article stands out for its comprehensive meta-analysis, []”). *Style differences*, such as using the word “information” rather than “abstract” in phrases like the following, “Based on the provided information, I would score this article as follows”. *Hypothetical justifications* (e.g., “However, I wouldn't score it a 4 because: *

The article doesn't seem to be introducing a new or innovative methodology or approach that significantly advances the field.”).

- **Magistral Small**: *Recapping the problem* (e.g., “Based on the provided abstracts and my understanding of academic quality standards, here is a detailed evaluation of the article []”). *Style differences*, such as starting with “### Evaluation of”. Incorrectly interpreting the few-shot examples (e.g., “The scores for the previous articles were not provided for this specific evaluation, but they would be similarly assessed based on their respective originality, significance, and rigour. [] Summary of Scores - **Article 1**: Score 1* - **Article 2**: Score 2* - **Article 3**: Score 3* - **Article 4**: Score 4*”).

Thus, some but not all models’ few-shot reports explicitly reflected the few-shot task by recapping it or making direct comparisons with the few-shot articles. Other models (e.g., Gemma3 27b) did not seem to mention the few-shot example at all. All sets of few-shot reports from a single LLM had some systematic differences with the zero-shot counterparts, however. Most notably, the style of the Gemma3 27b reports with few-shot were different from those without despite a lack of explicit acknowledgements of the examples provided. These changes were widespread (e.g., 98% of zero-shot and 49% of few-shot reports contained “evaluation”) and seemed to be entirely unrelated to few-shot, suggesting that the additional information prodded the model into being more likely to adopt different language styles in its report.

The zero-shot reports also had some features that were much rarer than in few-shot reports for all models. Most of the differences seemed to be stylistic but the DeepSeek 32b zero-shot reports contained more speculative language (e.g., “wonder” was in 26% of zero-shot and 17% of few-shot reports), Qwen 32b zero-shot reports contained more evaluative terms (e.g., “criterion”, “strengthen”, “demonstrates”), and Magistral Small zero-shot reports were much more likely to finish with a constructive criticism section (16% vs. 4% for phrases like, “**Constructive criticism:** To improve the rigour of this work, the authors should…”). A strange large stylistic difference for Llama4 Scout was that 61% of zero-shot reports contained “internationally” compared to 0.4% for few-shot. This originated from spelling out scores, as in “**Score: 3* (Internationally Excellent)**”.

In conclusion, some but not all models frequently made explicit reference to the few-shot examples, but these examples always resulted in systematic differences in the style of the reports.

### *Comparison of zero-shot model reports against each other*

Comparing the non-reasoning models against each other for zero-shot, the main differences were that Gemma3 27b tended to use a standard structure for the report, such as 83% of the time starting with numbered evaluations of originality, significance and rigour (e.g., “**1. Originality (3*)** The study demonstrates good originality.”), and giving explicit praise (e.g., “This article [] represents a substantial and valuable contribution...”) and criticism (e.g., “The abstract doesn’t clearly articulate *how* this study builds upon or fundamentally diverges from existing literature in this area.”). In contrast, Llama4 Scout explicitly mentioned “strengths” and “weaknesses” much more often than did Gemma3 27b.

Comparing the reasoning models against each other for zero-shot, the largest difference was that Magistral Small did not include a reasoning section. These reasoning sections tended to use informal, thoughtful and tentative language, although with different styles. For example, DeepSeek R1 32b used “I’m”, “trying”, and “think” much more than Qwen3 32 (61% vs. 20%, e.g., “That immediately makes me think this work could be original because...”). Another small difference is that DeepSeek R1 32b was more likely to use singular personal pronouns (e.g., “I’m”: 42% vs 5%) and Qwen3 32b was more likely to use plural personal pronouns (e.g., “let’s”: 71% vs. 16%, often in “Okay, let's start by…”).

Comparing the reasoning models against Gemma3 as a sample non-reasoning model, DeepSeek R1 32b was more likely to finish with a summative statement (e.g., “Putting it all together,…”), and to use first person terms like “I” and “me”. It also seemed to take a wider perspective in its thinking section by referring the “academic journal article” that it was evaluating (e.g., “journal” occurred in 84% of DeepSeek reports and 0.2% of Gemma3 reports). Qwen3 32b was also more informal and with much greater use of personal pronouns, compared to Gemma3 27b. Finally, Magistral favoured some task-relevant terms or concepts compared to Gemma3 26b, such as “innovative” (86% vs. 25%) and “theory-building” (34% vs. 0.1%).

These analyses illustrate that the report styles of these similar sized LLMs are very different for the same task. This may be due to their training data rather than their architectures.

### *Weighted sum correlations*

In theory, better results might be obtained by combining multiple models than by selecting the single best one. I fact higher correlations with the individual gold standard could be obtained by creating weighted sums of the individual model scores, but the difference was not large (Table 2). Its effect is like taking the mean rank of each article across all models. Nevertheless, a simple mean of all scores (or ranks) outperforms the best single Spearman correlation in all cases, so averaging different model scores (or ranks) is also a good strategy. It is also safer than identifying the optimal weighted sum statistically, because, in the 10-fold cross-validation tests, some folds produced much lower correlations. Moreover, simple means can be calculated without a gold standard whereas a gold standard is needed to identify the weightings for the weighed sums. For the secondary gold standard, the advantage of simple or weighted averages is less clear, however (Table 3).

Table 2. Weighted sum (of the 23 model scores, including few-shot variants) correlations (average of 10) and baseline correlations with the individual article gold standard. The mean/median fusion versions take the average LLM score for each article. The highest score is bold.

| UoA | Mean fusion Spearman | Median fusion Spearman | Best single Spearman | Rank average Spearman | Mean 10-fold Spearman |
|---|---|---|---|---|---|
| 1 | 0.387 | 0.389 | 0.364 | 0.391 | **0.403** |
| 2 | 0.419 | 0.395 | 0.408 | **0.422** | 0.420 |
| 3 | 0.489 | 0.463 | 0.461 | 0.490 | **0.494** |
| 4 | 0.515 | 0.491 | 0.488 | 0.512 | **0.519** |
| 5 | 0.456 | 0.418 | 0.408 | 0.467 | **0.467** |
| 6 | 0.547 | 0.491 | 0.510 | 0.549 | 0.525 |
| **All** | 0.463 | 0.442 | 0.427 | 0.465 | **0.497** |

Table 3. Weighted sum (of the 23 model scores, including few-shot variants) correlations (average of 10) and baseline correlations with the mean departmental score gold standard. The mean/median fusion versions take the average LLM score for each article. The highest score is bold.

| UoA | Mean fusion Spearman | Median fusion Spearman | Best single Spearman | Rank average Spearman | Mean 10-fold Spearman |
|---|---|---|---|---|---|
| 1 | 0.151 | **0.168** | 0.167 | 0.157 | 0.118 |
| 2 | 0.291 | 0.288 | **0.303** | 0.300 | 0.276 |
| 3 | 0.395 | 0.379 | 0.380 | 0.399 | **0.410** |
| 4 | 0.531 | 0.484 | 0.541 | 0.535 | **0.552** |
| 5 | 0.309 | 0.263 | **0.321** | 0.315 | 0.319 |
| 6 | 0.327 | 0.299 | **0.341** | 0.336 | 0.310 |
| **All** | 0.369 | 0.347 | 0.339 | 0.374 | **0.400** |

## Conclusions

The results show that a range of different open weights LLMs have an ability to score journal articles for research quality, including reasoning models, and that this ability may be present even for relatively small 4 billion parameter models in most UoAs, with 12 billion parameters being a safer minimum that worked for all six UoAs here. The results also give strong evidence that averaging scores from multiple identical prompts improves the results and weak evidence that few-shot prompting can help, although not necessarily by learning from the examples. It is possible that prompt variety is more important than few-shot examples when scores are averaged across multiple iterations. These results will not necessarily hold for models not tested here. In addition, alternative few-shot strategies may be more fruitful than those tried here.

These findings help to make LLM scoring of journal articles more practical through identifying smaller models that can work when there are limited computing resources and when the scoring is too sensitive to be exposed to public APIs, however secure. They also make this approach more credible in the sense that a range of different LLM sizes and architectures possess a research quality rating capability, so it seems more likely to be a core LLM ability rather than an artefact of specific architectures. An important caveat is evident in the violin plots: that even with few-shot LLM scores do not match the human scale and tend to avoid values much lower than 3*. Moreover, the violin plots also confirm that even average LLM scores for individual articles are unreliable. The main value of the LLM approach is therefore to rank articles to support decision making by showing that a given article is better than another, or above/below a given percentile threshold (e.g., top 5%). For this task, smaller models seem to be nearly as effective as larger ones, making their use practical in situations where high performance computing is not available, and data security is essential. Reasoning is not recommended because it substantially increases computing time/cost without improving the results.

In practical terms, deciding when to use LLM score predictions is a complex issue, involving considerations of systemic and ethical issues (Thelwall, 2025d) as well as the informational value of the scores. When making this decision, it would be useful to compare the level of agreement (or correlation) between human experts as well as between LLMs and expert scores. Unfortunately, there are no large-scale multidisciplinary datasets of post publication quality scores from multiple experts to identify the expected expert agreement rate, although analysis of reviewer data shows that expert disagreement is common (Bornman et al., 2010). Instead, as with citation-based indicators, they should be recognised as imperfect, treated with caution that is inversely proportional to their correlation with expert judgement, and (normally) used to support expert judgement, especially when it is uncertain (Thelwall, 2025de). In this context, the violin plots above may help to illustrate that the LLM score ranks are helpful, but not definitive.

# Declarations

**Funding and/or Conflicts of interests/Competing interests**: This study is partly funded by the Economic and Social Research Council (ESRC), UK (APP43146). The first author is a member of the editorial board of Scientometrics.

## Appendix: A Qwen3 32b short response example

<think> Okay, let's start by reading through the article and the user's query. The article is about RECQL4 in gastric cancer and its role in cisplatin resistance. The user wants me to score it as 1* to 4* based on originality, significance, and rigour. First, I need to break down each criterion. Originality would look for new findings, innovative methods, or novel insights. The study mentions a novel role for RECQL4 in regulating MDR1 via YB1. That seems original. They found a new signaling pathway, which is a plus. Significance is about influence on the field or policy. The study links RECQL4 to poor prognosis and drug resistance, which is important for gastric cancer treatment. If targeting RECQL4 improves therapy, that's significant. But I should check if this is a major breakthrough or just incremental. Rigour involves methodology and analysis. The study uses multiple patient samples, in vitro models, and shows both overexpression and silencing effects. That's good. They mention correlation with survival and functional assays. But are there any limitations? Maybe in vivo models would add more rigour, but the study is in vitro. Also, the mechanism is established through interactions and pathway analysis, which is solid. Looking at the 4* criteria: world-leading originality, significance, and rigour. The study does present a novel pathway but I'm not sure if it's at the highest level. The methods are robust, but maybe lacking some in vivo validation. Significance is high for therapeutic targets, but does it have broad impact beyond gastric cancer?

For 3*, internationally excellent but not top. The work is solid, contributes to understanding, and offers therapeutic rationale. The methods are thorough but perhaps not ground-breaking in technique. The significance is clear for gastric cancer treatment, which is a major area. So maybe 3*. I need to check if the study's originality is enough. They identified a new role for RECQL4 in transcriptional regulation via YB1-MDR1. That's a new mechanism. The significance is in improving treatment, which is important. Rigour is good with multiple experiments, but in vivo data could be missing. So overall, it's 3*. </think>

**Score: 3***

**Reasoning:**

### **Originality (3*)** The study identifies a novel mechanistic role [followed by one A4 page of the main report]

## Appendix: B Qwen3 32b short few-shot response example

<think> Okay, let's try to figure out how to score this article on the 1* to 4* scale based on originality, significance, and rigour. **The user provided examples for each score, so I need to compare this article to those examples**. First, originality. The article introduces synthetic light-curable polymeric materials that support dental pulp stem cells. They used a polymer microarray to screen materials and found triacrylates effective. The thiol-ene chemistry is highlighted as a method for rapid curing and low residual monomer. **Comparing to the previous examples, the 4* article discovered a new genetic cause for a disorder, which is a major original contribution**. This article is developing new materials, which is original, but maybe not as groundbreaking as a new genetic discovery. However, the use of thiol-ene chemistry in dental materials is novel. So maybe originality is strong, but perhaps not world-leading. Maybe 3* or 4*? Significance. The article addresses a need in regenerative dentistry by improving restorative materials. If these materials can support stem cells and lead to better regeneration, that's significant. The 4* article had a clear novel cause for a disease, which is very significant. This article's significance is in advancing dental materials, which could lead to clinical applications. The impact on clinical practice is notable, so maybe 3* or 4*. **But how does it compare to the 4* example?** The 4* example provided a new etiological understanding, which is foundational. This article is more about application, so maybe 3* for significance. Rigour. The methods include polymer microarray screening, thiol-ene chemistry, in vitro tests on DPSCs, and mechanical property comparisons. The experiments are detailed, and they compared different monomers. **The 4* article used whole-exome sequencing and functional studies, which are rigorous.** This article's methods are solid, but perhaps not as complex. The in vitro data is good, but in vivo testing is missing. Rigour is strong but maybe not at the highest level. So maybe 3*. Putting it together: Originality is strong (3*), significance is high (3*), rigour is good (3*). So overall 3*. **But let me check the examples again.** The 3* article on serotonin receptors used both clinical and preclinical studies, which is similar in combining different methods. The 4* article had a clear novel genetic finding with functional validation. This article's contribution is in material science for dentistry, which is important but maybe not as transformative. So 3* seems appropriate. However, the use of a polymer microarray to rapidly screen materials is innovative, which might push originality to 4*. But the 4* example had a more fundamental discovery. I think 3* is right here. The article is internationally excellent but not world-leading. </think>

**Score: 3***

**Reasoning:**

**Originality (3*):** The article demonstrates **international excellence** [followed by one A4 page of the main report]